\documentclass[usenatbib,onecolumn,useAMS]{mn2e}
\usepackage{graphicx}
\usepackage{amsmath}
\usepackage{amssymb}
\begin{document}
\voffset -2cm

\title[Halo Model and Environment]{A Halo Model with Environment Dependence: Theoretical Considerations}

\author[Gil-Mar\'in et al.]{H\'ector Gil-Mar\'in$^{1}$\thanks{Email: gil@ieec.uab.es}, Raul Jimenez$^{2}$\thanks{Email: raul.jimenez@icc.ub.edu}, Licia Verde$^{2}$\thanks{Email: liciaverde@icc.ub.edu} \\
$^1$ Institute of Space Sciences (IEEC-CSIC), Faculty of Science, Campus UAB, Bellaterra, Spain\\
$^2$ ICREA \& Institute of Sciences of the Cosmos (ICC), University of Barcelona, Barcelona 08024, Spain}

\maketitle

\begin{abstract}
We present a modification of the standard halo model with the goal of  providing an improved description of galaxy clustering.  Recent surveys, like the Sloan Digital Sky Survey (SDSS) and  the Anglo-Australian Two-degree  survey (2dF), have shown that there seems to be a correlation between the clustering of galaxies and their properties such as metallicity and star formation rate, which are believed to be environment-dependent. This environmental dependence  is not included in the standard halo model where the host halo mass is the only variable specifying galaxy properties. 
In our approach,  the halo properties i.e., the concentration, and the Halo Occupation Distribution --HOD-- prescription,  will not only depend on the halo mass (like in the standard halo model) but also  on the halo environment. We examine how different  environmental dependence of halo concentration and HOD prescription 
 affect the correlation function. 
We see that at the level of dark matter, the concentration of haloes affects moderately the dark matter correlation function only at small scales.
However  the galaxy correlation function is extremely  sensitive to the HOD details,  even when only the HOD of a small fraction of haloes is modified. 
\end{abstract}



\section{Introduction}

The modern language to describe analytically the clustering of galaxies is the halo model (e.g., \citet{cooray_sheth} for a review). In its original formulation, the halo model describes non-linear clustering of dark matter and can be applied to describe also the clustering properties of galaxies. 

 In particular, the halo-model can  be calibrated  to describe  galaxy clustering properties  in several different ways,  depending on how the observed properties of galaxies are to be related to the underlying dark matter  halo -- the so-called halo occupation distribution (HOD): using galaxy  abundance, their spatial distribution via the two-point correlation function or the luminosity dependence as function of the halo mass. In the halo occupation distribution, it is  customary  to classify galaxies as central or satellite. These two classes of galaxies are then assigned different occupation numbers  in the dark haloes.  In this description, all observational properties are completely specified by the dark matter halo mass. The original halo model has been remarkably successful at describing the first moment statistics of the clustering of galaxies. 

In reality, however, galaxies are not easily divided in central or satellite, and  the physical characteristics of galaxies of each type cannot be determined solely by the mass of the host halo: galaxies are more complex systems.  Environment must play an important role in the process of galaxy formation, the most 
striking observational evidence being that clusters today have a much higher fraction of early-type galaxies than is found in the field. It has been known for more than three decades that there is a relation between galaxy morphology and  density of the local environment starting from the results of  \cite{DavisGeller76, Dressler80,PostmanGeller84}, until most recent results e.g., \cite{Hogg02,Zehavi10}. It is not clear if this effect can be completely ascribed to the fact that the most massive haloes are naturally found in overdense regions, and that most massive haloes form on average earlier,  or if there is some extra environmental dependence; i.e., physical mechanisms such as ram-pressure stripping, harassment etc. \citep{gunn_gott, Moore96,Moore98} shape the properties of galaxies and operate in dense environments.
We know that, although {\it on average} the star formation and metallicity history are determined by the mass, their correlation properties are not \citep{SJPH,Mateus}. In particular, the clustering of the properties of galaxies has been shown to depend on more parameters than mass:
large-scale tidal fields have been shown to alter galactic spins  \citep{spins} and ellipticity \citep{Mandelbaum} but can also alter galaxy clustering properties \citep{Hiratatidal1,Hiratatidal2}. 

The fact that clustering properties of galaxies depend on galaxy internal properties, can be understood if dark matter haloes with different properties  and formation histories, cluster differently and host different galaxy populations. Consider for example the so-called halo assembly bias; e.g. \cite{gao_springel_white} found that the amplitude of the correlation function depends on halo formation time, thus haloes assembled at high redshift are more clustered than those formed recently.
Models of galaxy clustering statistics make some simplifying assumptions; mainly that {\it i) }number and properties of galaxies populating a dark matter halo depend only on the mass of the host halo, and {\it ii)} clustering properties of dark matter haloes are only a function of their mass and not of the larger environment (this assumption is at the basis of the excursion set-formalism).

It is well known that the formation time of haloes depends on the halo mass and that objects formed at early times tend, on average,  to be more concentrated than objects that formed recently.   In the hierarchical galaxy formation model there is a correlation between galaxy type and environment induced by the fact that the mass function of  dark haloes in dense regions in the Universe is predicted to  be top-heavy. In a simple improvement of the classic halo model \citep{AbbasSheth1, AbbasSheth2}  the most massive haloes only populate the densest regions and a correlation between halo abundance and environment is introduced. This approach matches the prediction of the excursion set approach; all correlation with environment comes from mass.

But there are indications that there  could be a more complex dependence on the environment which is not fully described by halo mass e.g., \cite{ShethTormen2004} and discussion above. 
It has been known for a while that the star formation history is perhaps the most affected quantity by environment, this is  the so-called downsizing effect, e.g. \citep{cowie,heavens}. Furthermore, the recent zCOSMOS survey has shown how other variables are also affected by environment, in particular the shape of galaxy stellar mass function \citep{zcosmos1,zcosmos2}, and the  formation of red galaxies at the low mass end \citep{zcosmos3,zcosmos4,galaz10}. All these new observational results seem to indicate that the physical properties of galaxies are, at least in part determined by the environment in which they live,  and that this dependence goes beyond the fact that the halo mass function is influenced by the local background density. Therefore, analyses of future surveys with a similar sensitivity as the zCOSMOS and, especially, large redshift coverage, may benefit  from  theoretical tools that can include  environmental dependence. In practice the definition of ``environment" may not be easy or univocal, and the definition may change as a function of the galaxy property under consideration. Here we simply set up the mathematical description assuming that ``environment" has been  already defined.
We extend the standard halo model  to include an extra dependence on the environment,  classified as  ``cluster/node", ``filament" and ``void" although the treatment  is general enough that other choices or interpretations of environment are possible. Motivated by the excursion-set model  expectations, the extra dependence comes through the halo density profile and mass function for the dark matter and also through the HOD parameters for galaxies. In the  excursion-set model  this is given by the formation time but in our model we introduce an extra parameter set by the environment, leaving us one more degree of freedom to tune the HOD. As a starting point, here we introduce the formalism and show the effect of each of the parameters of the model on the correlation function.  However  we expect that the full potential of the model and the relevance of its features  will appear when considering   clustering statistics beyond the  2-point function of the over-density field such as e.g. the marked correlation functions \citep{Skibba06,SJPH,Mateus}.   We defer this to forthcoming work \cite{Hectorinprep}. 
  
This paper is organised as follow: in \S 2 we present the theoretical formulation of our extension of the model. We first introduce all the model parameters and ingredients, then in \S \ref{section24} we present the expression for the correlation function of dark matter and galaxies. In \S 3 we analyse how an scenario environments with different properties can affect to the correlation function. In \S 4 we present the conclusions of our work. Accordingly in the Appendices we review the basics of the standard halo model and the halo occupation distribution and give details of equations presented in \S 2.

\section{The Model}

The halo model provides a physically-motivated way to estimate the two-point (and higher-order) correlation function of dark matter and galaxy density field. We review the classic halo model in Appendix A. Despite its simplicity, the model is extremely successful: the model's predictions have  been compared with both simulations and observations, reproducing well the clustering properties.
Nevertheless,  it presents some limitations and shortcomings which we discuss next. The background material presented in Appendix A is useful to set up the stage of motivating our extension of the model and to define symbols and nomenclature used. We thus refer the reader to Appendix A for definitions of many of the symbols used.

In the halo model every halo is characterised only by its mass. This is correct at the leading order, and has been so far sufficient. Much more accurate measurements of dark matter and galaxies clustering will be available in the near future (e.g., SDSSIII, EUCLID, JDEM, DES, BigBOSS\footnote{See http://bigboss.lbl.gov/index.html and http://arxiv.org/abs/0904.0468 for mote details}, PannStarr, LSST\footnote{See http://www.lsst.org/lsst for more details}, etc.) and a more sophisticated modelling may be needed to describe such data-sets. In fact there are indications already that the properties of galaxies depend somewhat on the environment, and not only on the host halo mass. See \S 1. 

It is clear that the mass has to be the primary variable in any halo model formalism. However in order to include some environmental dependence  another variable needs to be introduced. At the level of the galaxy correlation function this ``variable" can be modulating  the HOD: in some environments galaxies can populate haloes in a different way than in other. At the level of dark matter we can introduce an environmental dependence through the concentration of the density profile. It has been observed that the relation between the mass and the concentration has a very high dispersion: the distribution
around the mean concentration is approximately independent of halo mass, and is well approximated by a lognormal with $rms$ $\sigma_{ln c}=0.3$ (e.g., \cite{ShethTormen2004}).  In the halo model formulation a relation between concentration $c$ and mass $m$ is assumed (see Eq. \ref{concentration}), however this high dispersion may indicate that the concentration depends on extra (hidden) variables,  such as halo formation history (haloes which form at high redshift  have a higher concentration), tidal forces etc.,  that in the end would depend on the environment.  A possible approach to this problem was presented by \cite{giocoli10}. It consists in treating the concentration as a stochastic variable and not only integrate over the mass but over all possible values of the concentration. In this work we will instead consider  that the relation between concentration and mass may depend on the environment.

In our model, we assume that the Universe contains three kind of structures or environments: nodes, filaments and voids. Halos lie either in a node or filament regions. Voids  contain no haloes but occupy a large fraction of the Universe's volume, especially at $z\simeq0$.

At the level of dark matter correlation function, we  distinguish between haloes in the node regions (hereafter node-like haloes) and haloes in the filament regions (hereafter filament-like haloes) through the concentration of their profile. We assume that the node-like haloes have a  profile with a concentration function $c_{nod}(m)$ and  filament-like haloes with  $c_{fil}(m)$. For simplicity, in presenting our equations and our figures,  we assume  that these concentrations are constants, but the formalism can straightforwardly be generalised to allow  these concentrations to be  functions of the mass, as it is done  in the halo model. The range of these values is expected to be between $\sim 1$ and $\sim 15$;  these values (or functions) could be calibrated by N-body simulations.

At the level of galaxy correlation function, we can decide to populate haloes in  different ways, according to the type region where they lie. For instance, node-like haloes may have less satellite galaxies, or more likely to have a central galaxy than filament-like haloes. 

In this section we show how we can modify mathematically the halo model in order to take into account all these effects.

\subsection{Mass function} \label{mass_function_section}

As in the standard halo model (see Appendix A for details), the halo number density is given by the mass function, which  can be computed analytically in the extended Press Schechter approach or calibrated on N-body simulations. For this application we will adopt the  \citet{st} mass function.
More massive haloes are more likely to be found in high-density environments, thus in principle the mass function can be made to be dependent on the environment and the dependence could be calibrated on N-body simulations. At this stage, this procedure is similar to the approach presented by  \cite{AbbasSheth1, AbbasSheth2}. For this application the environment-dependence of the mass function is accounted for through the  fact that the mass function depends on the mean matter density: $\bar\rho_m$, which we consider to be dependent on the local environment. However, this model is general enough to use any mass function with any environmental dependence.
Let us consider a region $i$ with a volume $V_i$ and with a mean matter density $\bar \rho_{m_i}$. If we consider this region of the Universe alone, then the number of objects of mass $m$ in this $i$ regions per unit of volume (of this region) and per unit of mass, namely $n_i(m)$, is related to the mass function of the whole Universe as \citep{AbbasSheth1}, 
\begin{equation}
n_i(m)=\left[1+b(m)({\cal Y}_i-1)\right]n(m)
\label{mass_function1}
\end{equation}
where ${\cal Y}_i$ is the ratio between the mean matter densities of the region $i$ and of the whole Universe: ${\cal Y}_i\equiv\bar\rho_{m_i}/\bar\rho_m$. It is also useful to define the volume fraction of this region $i$, $V_i$ and the volume of the whole (observable) Universe $V$ as ${\cal X}_i\equiv V_i/V$.
Conservation of mass, imposes that the consistency relation must be satisfied: 
\begin{equation}
\sum_i{\cal X}_i {\cal Y}_i=1
\end{equation}
where the summation runs over all kinds of structures, in our case nodes, filaments and voids. Since we assume that there are no haloes  in voids (${\cal Y}_v=0$), this equation reads: ${\cal X}_n {\cal Y}_n + {\cal X}_f {\cal Y}_f=1$.\footnote{Here the sub-indices $n$, $f$ and $v$ accounts for the regions nodes, filaments and voids respectively} 
In other words, since the mass function is proportional to the mean matter density (see Eq. \ref{mass_function}), if we consider a region with a mean matter density ${\cal Y}_i$ times denser than the mean density of the Universe, then the mass function of this region will be ${\cal Y}_i$ times the one of the whole Universe (see Eq. \ref{mass_function1})

For illustration,  here the adopted values for these parameters are  obtained from \cite{aragon-calvo} and are listed in table \ref{volume_table}\footnote{In their work, the authors split their haloes in four types: clusters, filaments, walls and voids. Here we adopt the node values of ${\cal X}$ and ${\cal Y}$ as their clusters haloes. For the voids we adopt ${\cal X}$ but we set ${\cal Y}$ to 0, and we combine their filaments and walls values of ${\cal X}$ and ${\cal Y}$ into what we call filaments.}. In this work we choose these numbers as a fiducial values, but  other values are possible. In fact, the values of these parameters will depend mainly on the definition of environment, and on the definition of what a node and a filament is.

\begin{table}
\begin{center}
\begin{tabular}{cccc}
 & $\cal X$ & $\cal Y$ & ${\cal X}\cdot{\cal Y}$\\ 
\hline
\hline
Nodes & $3.8\times10^{-3}$ & 73 & 0.2774\\
Filaments & 0.1368 & 5.28 & 0.7223 \\
Voids & $\sim 0.86$ & 0 & 0\\
\hline 
\end{tabular}
\end{center}

\caption{Volume fraction ${\cal X}$ and mean density fraction ${\cal Y}$. Values for node-,  filament-like regions and voids used in this work. Data from \citep{aragon-calvo}}
\label{volume_table}
\end{table}

The filaments regions are more abundant than the node regions, but  less dense. The voids occupy almost $90\%$ of the volume.

\subsection{Halo density profile}

In the halo model, the halo density profile can be written as a function of mass and  concentration: $\rho(r|m,c)$. Usually  the NFW profile \citep{nfw96,nfw97} is adopted, with an empirical relation between the concentration and the mass such as  Eq. \ref{concentration}.  
For this application we will consider that these two variables are independent and that the concentration is set by the environment of the halo. Thus, we have two different profiles depending on whether the halo lies in a node-like region, $\rho(r|m,c_{nod})$ or in a filament-like one, $\rho(r|m,c_{fil})$. In principle $c_{nod}$ and $c_{fil}$ can be  functions of the mass, which could be  calibrated  from N-body simulations. Here however, for simplicity,  we will  consider that these two variables are just constants. 

\subsection{Halo Occupation Distribution}
In our approach,  the  standard HOD (described in Appendix A \S \ref{section_hod_chm})  must also be modified to include a possible  dependence on environment. 

As an example see the discussion in \citet{zehavi05}: blue and red galaxies populate haloes of same mass in a different way (see Appendix A \S \ref{section_hod_chm} and Fig. \ref{hod_rb}); blue galaxies tend to occupy low-density regions while red galaxies the high-density ones. 

Here we are interested in  modelling generic properties of galaxies (star formation, metallicity etc.) not just colours. Whatever the property under consideration is,  in our modelling galaxies  can still be divided in two classes (nodes and filaments); for simplicity here we still call them  ``red" and ``blue", but one should bear in mind that the argument is much more general. 
Thus we make the HOD to depend not only on the mass of the  host halo but also on its environment (node or filament) by adopting two different prescriptions for nodes or filaments: the parameters $M_{min}$, $M_1$ and $\alpha$ of Eq. \ref{HOD_cen} and \ref{HOD_sat} must be specified for node- and filament-like haloes.  Let us continue with  the working example of red and blue galaxies.
While in the standard halo model, the red and blue galaxies are uniformly distributed inside haloes, in the extreme case of this environmental dependence,  blue galaxies only populate haloes which are in  filament regions while red galaxies only populate node-like haloes. However this is an extreme segregation and in a more realistic scenario  there will be some mixing: node-like haloes host some blue galaxies and filament-like haloes can host  some red ones.  In order to describe this, we introduce the segregation index ${\cal S}$. When ${\cal S}=0$, there is no segregation, i.e.  red and blue galaxies are distributed equally in filament and node haloes: this corresponds to the standard  approach with no environmental dependence. In the other limit, if ${\cal S}=1$, blue galaxies live in filament haloes and red ones in node haloes: this is the extreme case of our extended halo model with environmental dependence. 

Thus our adopted HOD equations are (see Appendix A for definition of HOD):
\begin{eqnarray}
g_{nod}^i(m)&=&\left(\frac{1+{\cal S}}{2}\right)g_{red}^i(m)+\frac{1-{\cal S}}{2}g_{blue}^i(m) \nonumber \\ 
g_{fil}^i(m)&=&\left(\frac{1+{\cal S}}{2}\right)g_{blue}^i(m)+\frac{1-{\cal S}}{2}g_{red}^i(m)\,.
\label{segregation}
\end{eqnarray}
where ${\cal S}$ is the segregation index, $g_{property}(m)$ denotes the average number of galaxies of a certain $property$ in a halo of mass $m$ and $g_j(m)$ denote the average number of galaxies in nodes ($j\rightarrow nod$) and in filaments ($j\rightarrow fil$). This $g(m)$ function is given by Eq. \ref{HOD_cen} and \ref{HOD_sat} for central ($i\rightarrow cen$) and satellite galaxies ($i\rightarrow sat$) respectively. 
These equations have 3 parameters: the minimum mass for a halo to host one (central) galaxy, $M_{min}$; the average mass for a halo to host its first satellite galaxy, $M_1$; the power-law slope of the satellite mean occupation function, $\alpha$. Taking into account that the HOD for red and blue galaxies may be different there are 6 HOD parameters. Moreover, if the two populations are not completely segregate but there is a degree of mixing, we have an extra parameter (see Appendix A for more details). In what follows, for some applications we will set $g_{nod}=g_{fil}$ (i.e. $g_{red}=g_{blue}$) but in general the HOD parameters for  $g_{red}$ will be different from those of $g_{blue}$ and thus, in general, $g_{nod} \ne g_{fil}$.

In next sections we will see how having two different HOD for nodes and filament regions modifies the correlation function respect to the case of having the same HOD for all galaxies.

\subsection{Two-point correlation function}\label{section24}
\begin{figure}
 \centering
\includegraphics[scale=0.25]{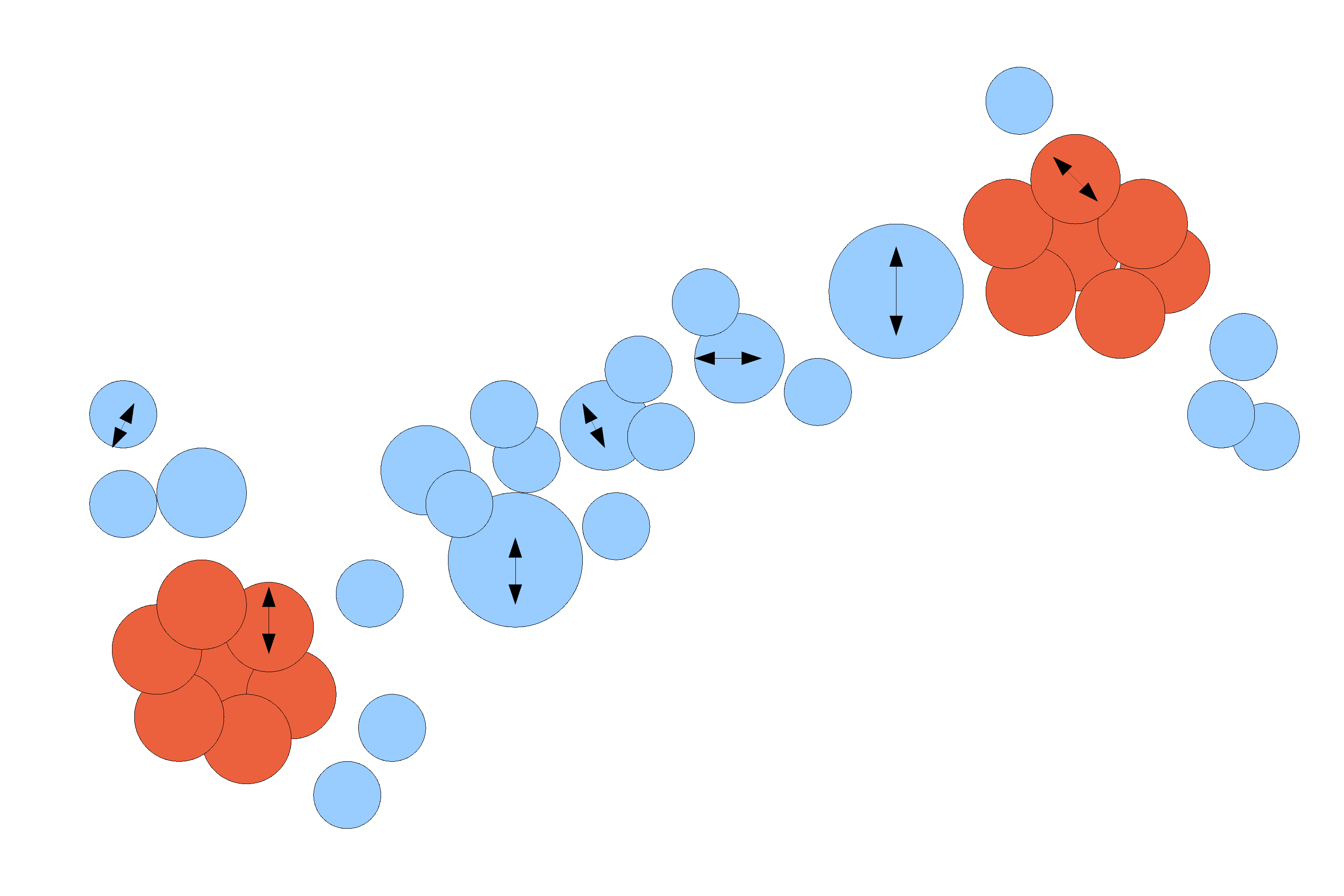}
\includegraphics[scale=0.25]{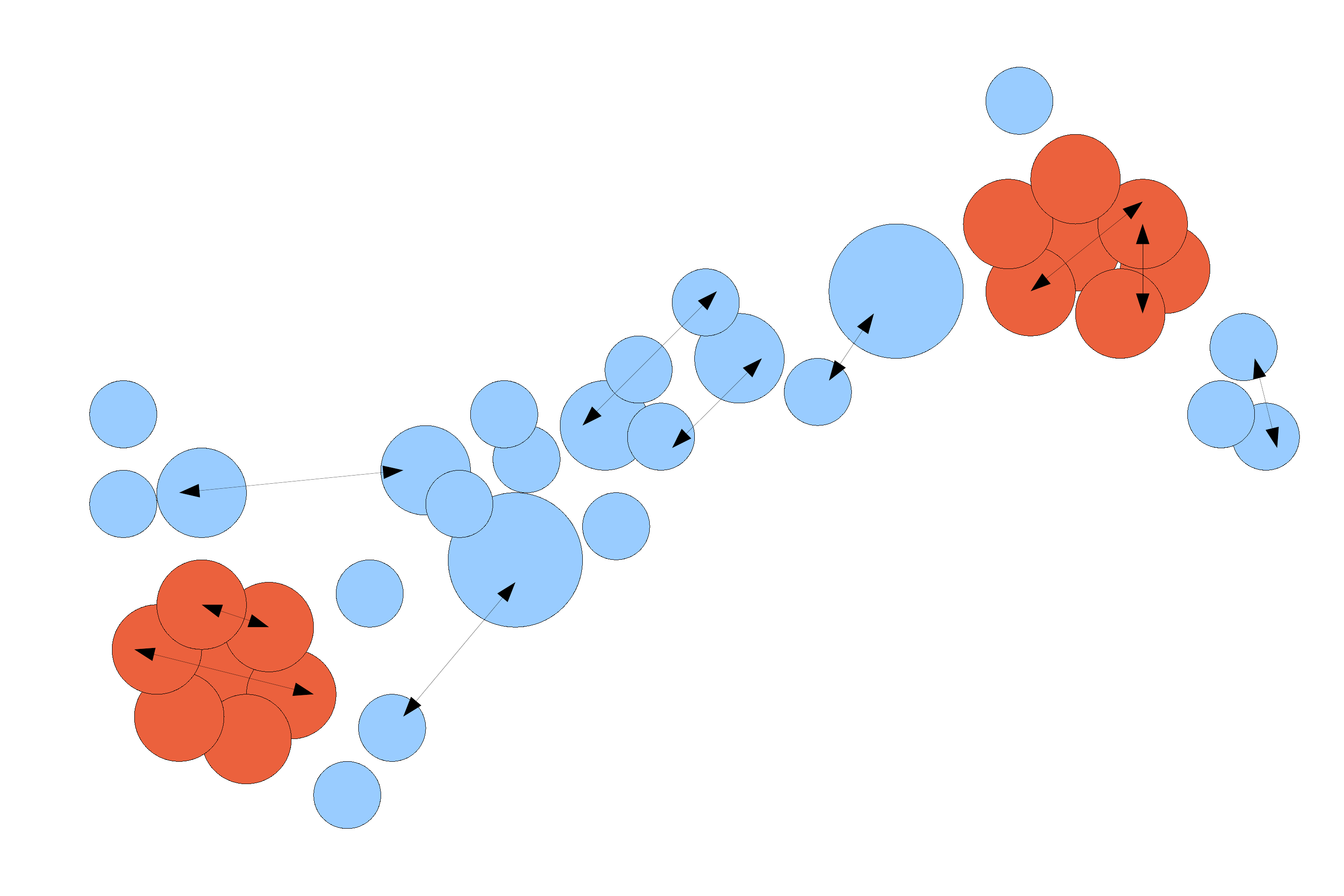}

\includegraphics[scale=0.25]{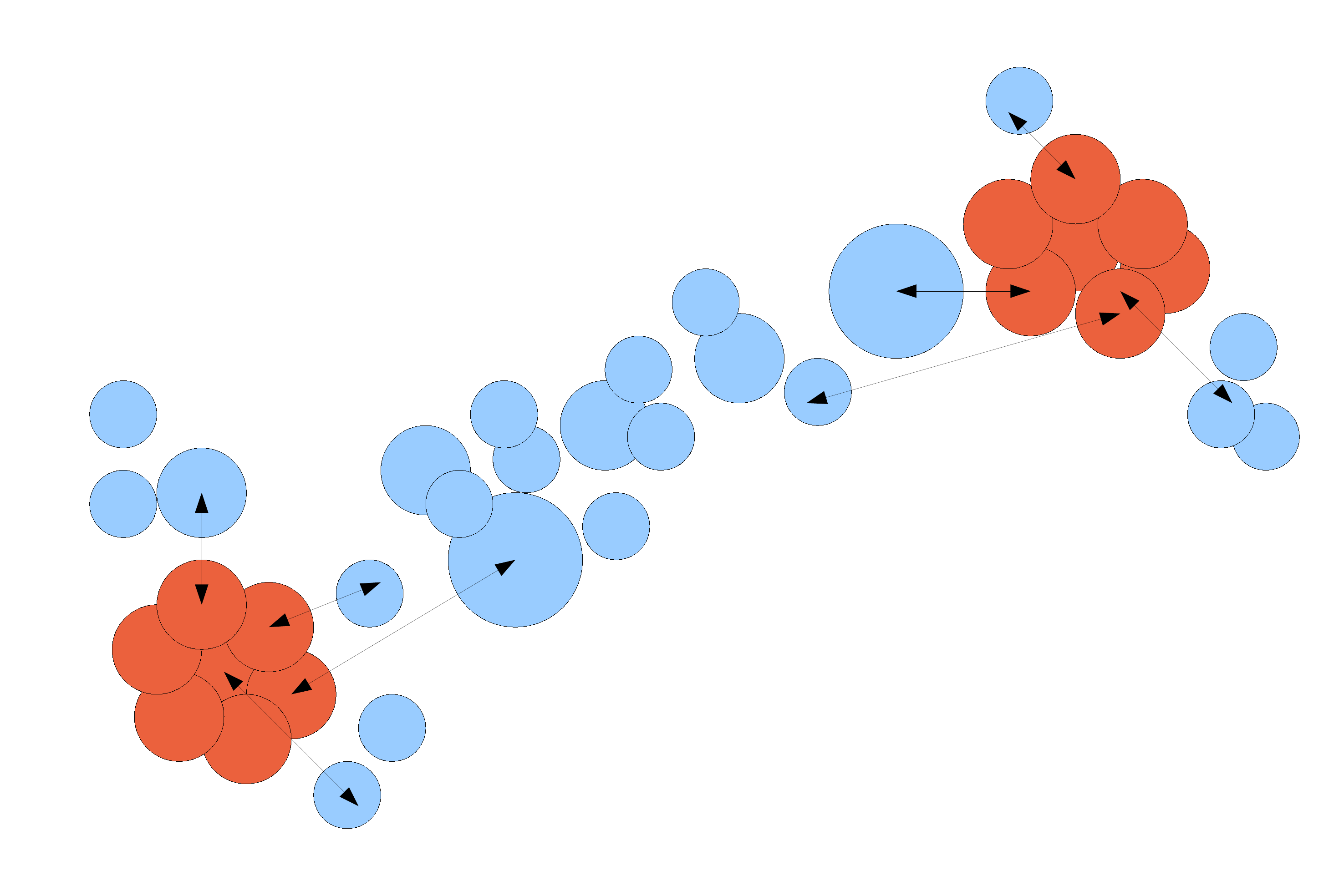}
\caption{In our modified halo model, dark matter haloes are of two kinds: node-like (red circles) and filament-like (blue circles). Because of this, our correlation function is split in three terms: the one-halo-one-$\eta$ term (top-left picture) describes the interaction of particles inside the same halo, the two-halo-one-$\eta$ (top-right picture) term involves particles which are in different haloes of the same kind. Finally the two-halo-two-$\eta$ term (bottom picture) considers the contribution of particles which are in different haloes of different kinds.}
\label{our_terms}
\end{figure}

Following the above  assumptions, we can compute the two-point correlation function for dark matter and galaxies. Since we have two types of structures (nodes and filaments), the correlation function is split in three terms instead of two as in the halo model:\footnote{The notation in this paper is the following: $1\eta$ means that both particles belong to the same kind of halo whereas $2\eta$ to different one.}
\begin{equation}
\xi({\bf r})=\xi^{1h1\eta}({\bf r})+\xi^{2h1\eta}({\bf r})+\xi^{2h2\eta}({\bf r})
\end{equation}
where now $\xi^{1h1\eta}({\bf r})$ accounts for particles in the same halo, $\xi^{2h1\eta}({\bf r})$ for particles in  different haloes but of the same kind and $\xi^{2h2\eta}({\bf r})$ for particles in  different haloes and of different kind separated by a distance ${\bf r}$. In Fig. \ref{our_terms} we show schematically the contribution of these different terms.

\subsubsection{Dark Matter correlation function}

For dark matter particles the correlation function reads,
\begin{equation}
\xi_{dm}({\bf r})=\xi_{dm}^{1h1\eta}({\bf r})+\xi_{dm}^{2h1\eta}({\bf r})+\xi_{dm}^{2h2\eta}({\bf r})
\label{ehm_dm_terms}
\end{equation}
where each terms is, (see Appendix B for derivation)
\begin{eqnarray}
\label{xidm1}\xi_{dm}^{1h1\eta}({\bf r})&=&\sum_{i=1}^2 \int dm\, {\cal X}_i \frac{m^2 n_i(m)}{\bar\rho_m^2}\int_{V_i}d^3{\bf x}\, u(x|m,c_i) u(|{\bf x+r}||m,c_i)\\
\label{xidm2}\nonumber\xi_{dm}^{2h1\eta}({\bf r})&=&\sum_{i=1}^2\int dm'\, {\cal X}_i \frac{m' n_i(m')}{\bar\rho_m}\int dm''\, {\cal X}_i \frac{m'' n_i(m'')}{\bar\rho_m}\int_{V_i}d^3{\bf x'}\, u(x'|m',c_i)\int_{V_i}d^3{\bf x''}\, u(x''|m'',c_i)\times\\
&\times&\xi_{hh}(|{\bf x'-x''+r}||m',m'')\\
\label{xidm3}\nonumber\xi_{dm}^{2h2\eta}({\bf r})&=&2\int dm'\, {\cal X}_1 \frac{m' n_1(m')}{\bar \rho_m}\int dm''\, {\cal X}_2 \frac{m'' n_2(m'')}{\bar\rho_m}\int_{V_1} d^3{\bf x'}\, u(x'|m',c_1) \int_{V_2} d^3{\bf x''}\,u(x''|m'',c_2)\times\\
&\times&\xi_{hh}(|{\bf x'-x''+r}||m',m'')\,,
\end{eqnarray}
 where $i=1$ refers to nodes and $i=2$ to filaments.
 Here $n(m)$ is the halo mass function, $u(x|m,c)$ is the normalised density profile of the halo of mass $m$ and concentration parameter $c$ at a radial distance $x$ (see Eq. \ref{nor_profile}), $\bar\rho_m$ is the mean matter density, $\xi_{hh}(d|m_1,m_2)$ is the two-point correlation function of two haloes of masses $m_1$ and $m_2$ separated by a distance $d$ (see Eq. \ref{xi_hh} for the model used here) and ${\cal X}_i$ and ${\cal Y}_i$ are the volume and density fractions defined in \S \ref{mass_function_section}. The integration $\int_V d^3\bf{x}$ runs over all haloes' volume and $\int dm$  runs over all halo mass range (see Appendix A for computational details). 
\begin{figure}
\centering
\includegraphics[scale=0.7]{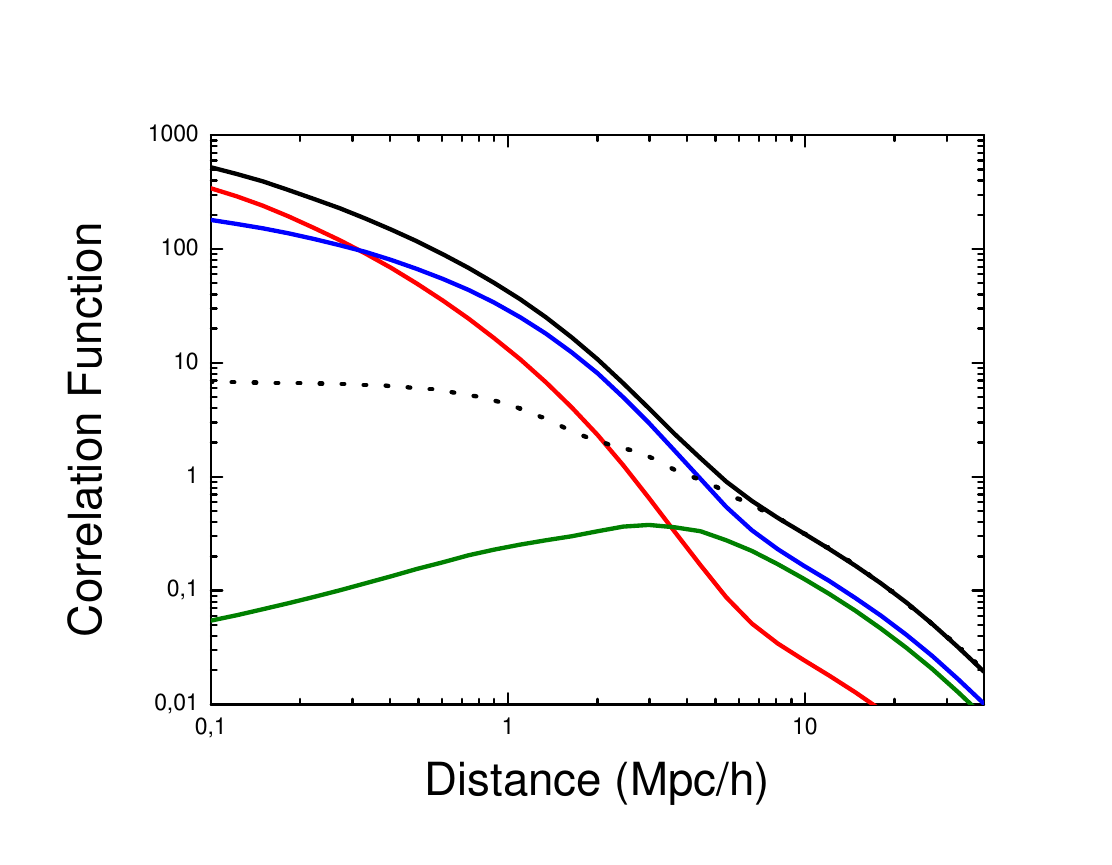} 
\includegraphics[scale=0.7]{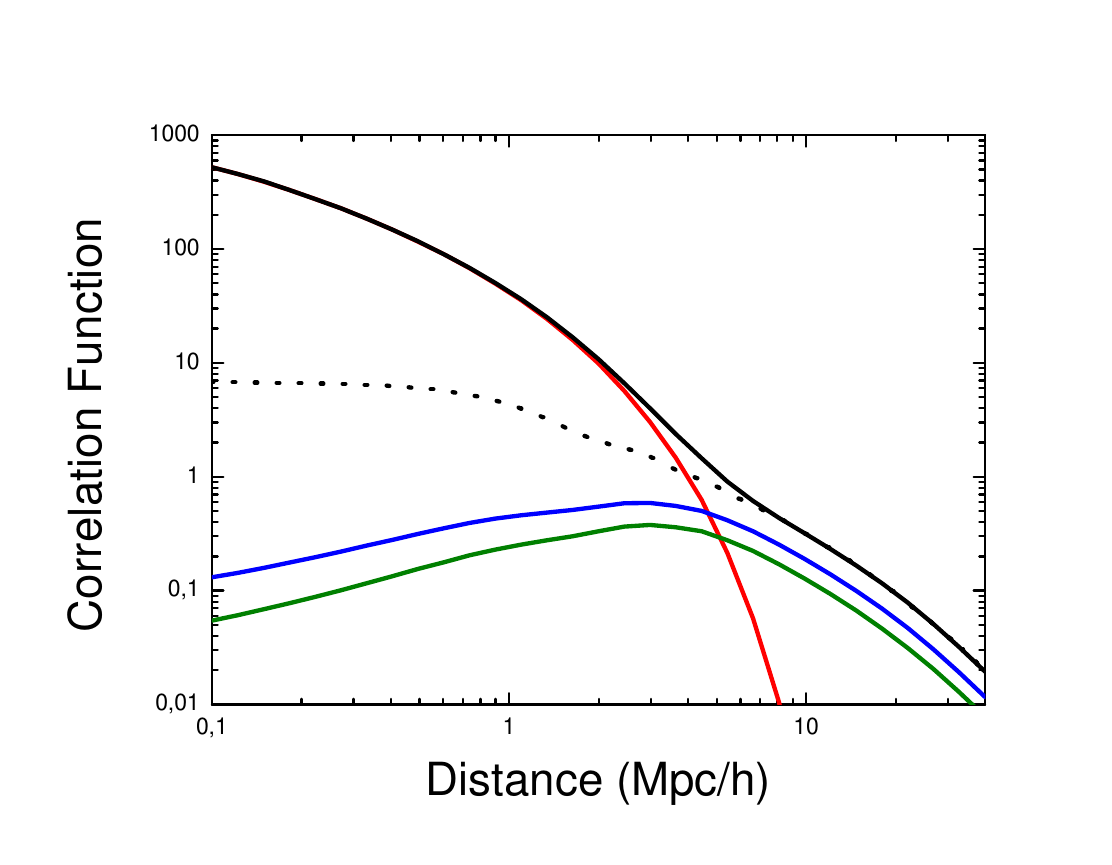}
\caption{Different contributions to the dark matter correlation function according to our model (Eqs. \ref{xidm1} - \ref{xidm3}). Left panel: total correlation function $\xi_{dm}$ (black solid line), node contribution $\xi_{dm}^{nod}$ (red line), filament contribution $\xi_{dm}^{fil}$ (blue line), cross contribution $\xi^{2h2\eta}_{dm}$ (green line) and linear dark matter contribution $\xi_{lin}$ (black-dotted line). Right panel: total correlation function $\xi_{dm}$ (black solid line), 1-halo term $\xi_{dm}^{1h1\eta}$ (red line), 2-halo-same-halo term $\xi_{dm}^{2h1\eta}$ (blue line), 2-halo-different-halo term $\xi_{dm}^{2h2\eta}$ (green line) and $\xi_{lin}$ (black-dotted line).}
\label{plot_concentration}
\end{figure}
In Fig.\ref{plot_concentration} we show the contribution of the different terms. Here we have assumed $c_{nod}=10$ and $c_{fil}=2$, just as an example  to introduce the environmental dependence. 
According to the  adopted values of ${\cal X}$ and ${\cal Y}$, the effect of the nodes is  dominant at small scales ($r<0.3$ Mpc/h) (red line in left panel). At intermediate scales ($0.3<r<6$ Mpc/h) the filaments dominate and at large scales ($r>6 Mpc/h$) both filaments and the cross term play an important role in the total correlation function (blue and green lines in left panel). This is expected due to the fact that the filaments are more abundant than nodes but nodes are more concentrate than filaments.

In the right panel we show the contribution of the different terms of Eq. \ref{ehm_dm_terms}. As we expected, the $1h1\eta$-term (red line) term dominates at sub-halo scales whereas $2h1\eta$-term (blue line) dominates at large scales. The cross correlation term between nodes and filaments, $2h2\eta$-term (green line) is  less important than filaments at large scales but is not negligible.
In the left panel, the shape of the filament and the node contribution is set by   the value of the concentration we have chosen. We will come back to this point in the next section to see how the concentration parameter affects the shape of the correlation function. In both panels the linear correlation function is also plotted (black-dotted line). The ratio between $\xi_{lin}$ and $\xi_{dm}$ at large scales can be defined as an effective large scale dark matter bias,
\begin{equation}
b_{dm}\equiv \int dm \frac{m n(m)}{\bar\rho_h}b(m)=1\,
\end{equation}
by definition of $b(m)$ (see Eq. \ref{bias}) this dark matter bias is one if we integrate over all range of masses.
Also the effective large scale bias can be defined for the node and filament contribution, this effective bias is given by
\begin{equation}
 b_{dm}^i\equiv{\cal X}_i \int dm \frac{m n_i(m)}{\bar\rho_m}b(m)={\cal X}_i{\cal Y}_i\,.
\label{dm_bias}
\end{equation}
Here $i$ stands for either nodes or filaments and the relation $b_{dm}=b_{dm}^{nod}+b_{dm}^{fil}=1$ is satisfied. 
Note that the only difference between the the dark matter bias of nodes and filaments is due to ${\cal X}_i$ and ${\cal Y}_i$. We will see that if we rescale the mean density $\bar\rho_m$ to the mean density of nodes or filaments (${\cal X}_i \bar\rho_{m_i}$), then the effective bias for nodes and filaments is the same.

Note that when we refer to  the node and filament contribution to the total correlation function, we refer to the terms of the sum of Eq. \ref{ehm_dm_terms}. These terms are different from the correlation function we would obtain if we only took into account the node- or filament-like haloes (we call it 'pure' node and filament contribution).  In the first case, each term is divided by the total dark halo-matter density squared, $\bar \rho_m^2$,  while in the second ('pure')  case instead  it would be divided by the density of node- or filament-like haloes.\footnote{ In order to obtain these `pure' node and filament terms we have to multiply the mean density $\bar\rho_m$ by the factor ${\cal X}_i {\cal Y}_i$, in order to obtain the mean density of nodes ($i=1$) or filaments ($i=2$)}
In Fig. \ref{plots_hods} (left panel) the contribution of these `pure' terms is shown for the dark matter. Note that the shape is the same as  in Fig. \ref{plot_concentration} and the only difference is that the normalisation  is shifted  by a factor $({\cal X}_i{\cal Y}_i)^{-2}$.

\subsubsection{Galaxy correlation function}
In this case the environmental dependence can be introduced not only through the concentration parameter of the host halo, but also through the HOD: galaxies do not populate in the same way filament- and node-like haloes even if the host halo has the same mass. In other words, if we have two galaxy populations (red and blue) with different HOD, e.g., red galaxies are more abundant in node haloes and blue in filament haloes, then an environmental dependence arises. This is the case presented here.

\begin{figure}

\centering
\includegraphics[scale=0.7]{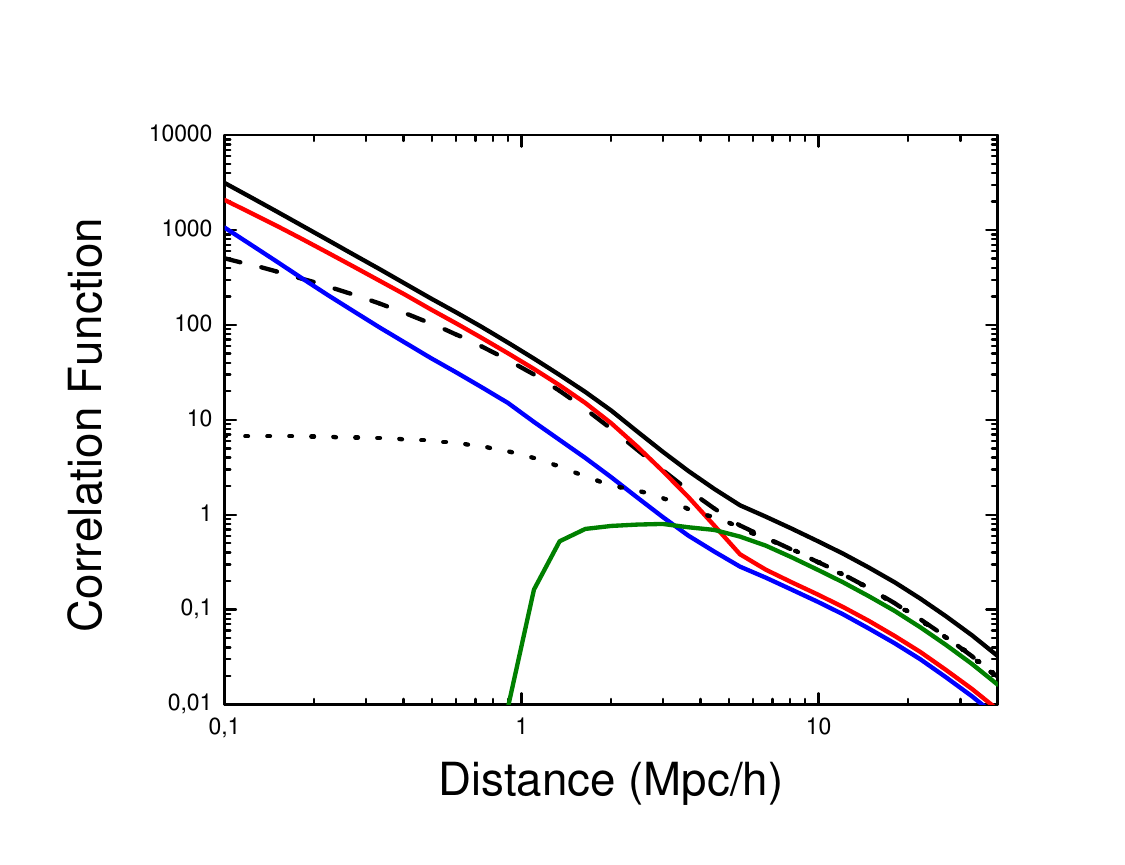}
\includegraphics[scale=0.7]{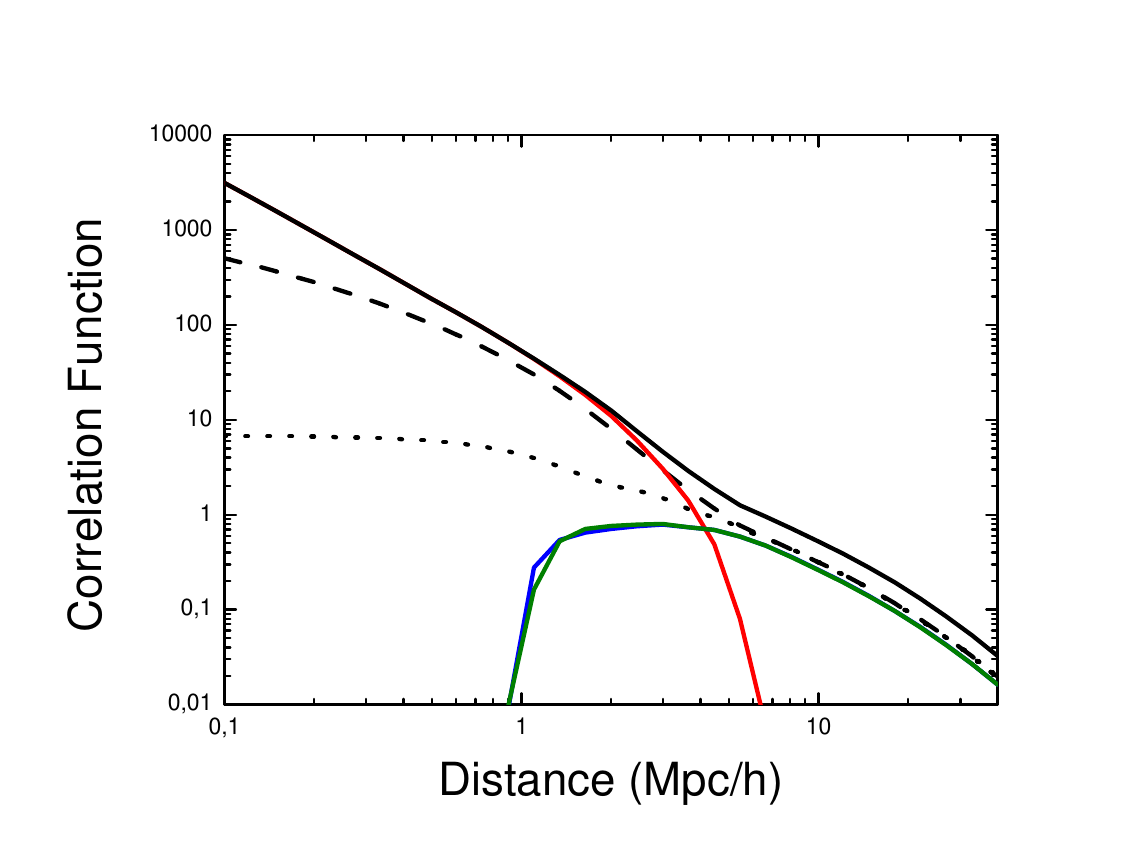}
\caption{Same notation as in Fig. \ref{plot_concentration}. Left panel: $\xi_{gal}$ (black solid line), $\xi_{gal}^{nod}$ (red line), $\xi_{gal}^{fil}$ (blue line) and $\xi_{gal}^{2h2\eta}$ (green line). Right plot: $\xi_{gal}$ (black solid line), $\xi_{gal}^{1h1\eta}$ (red line), $\xi_{gal}^{2h1\eta}$ (blue line) and $\xi_{gal}^{2h2\eta}$ (green line). In both panels $\xi_{dm}$ (black dashed line) and $\xi_{lin}$ (black dotted line). }
\label{fig7}
\end{figure}

As before, the galaxy correlation function reads,
\begin{equation}
\xi_{gal}({\bf r})=\xi_{gal}^{1h1\eta}({\bf r})+\xi_{gal}^{2h1\eta}({\bf r})+\xi_{gal}^{2h2\eta}({\bf r})
\label{ehm_gal_terms}
\end{equation}
where each term is,
\begin{eqnarray}
\label{xigal1}\xi_{gal}^{1h1\eta}({\bf r})&=&\sum_{i=1}^2\int dm\,{\cal X}_i  \frac{n_i(m)}{\bar n_{gal}^2}\left[2\,g_i^{cen}(m)g_i^{sat}(m)u(r|m,c_i)+\left(g_i^{sat}(m)\right)^2\int d^3{\bf x}\, u (x|m,c_i)\, u(|{\bf x + r}|m,c_i)\right]\\
\label{xigal2}\nonumber\xi_{gal}^{2h1\eta}({\bf r})&=&\sum_{i=1}^2\int dm'\, dm''\, {\cal X}_i^2 \frac{n_i(m')n_i(m'')}{\bar n_{gal}^2}\left[g_i^{cen}(m')g_i^{cen}(m'')\xi_{hh}( r|m',m'')+2\,g_i^{cen}(m'')g_i^{sat}(m')\right.\times\\
&\times&\int d^3{\bf x'}\, u(x'|m',c_i)\xi_{hh}(|{\bf x' + r}||m',m'')+g_i^{sat}(m')g_i^{sat}(m'')\int d^3{\bf x'} d^3{\bf x''}\,  u(x'|m',c_i) u(x''|m'',c_i)\\
\nonumber&\times&\left.\xi_{hh}(|{\bf x'-x'' + r}||m',m''\right]\\
\label{xigal3}\nonumber\xi_{gal}^{2h2\eta}({\bf r})&=&2\int dm'\, dm''\, {\cal X}_1 {\cal X}_2 \frac{n_1(m')n_2(m'')}{\bar n_{gal}^2}\left[g_1^{cen}(m')g_2^{cen}(m'')\xi_{hh}( r|m',m'')+\,g_1^{cen}(m')g_2^{sat}(m'')\right.\times\\
&\times&\int d^3{\bf x''}\, u(x''|m'',c_2)\xi_{hh}(|{\bf x'' + r}||m',m'')+\,g_2^{cen}(m'')g_1^{sat}(m')\int d^3{\bf x'}\, u(x'|m',c_1)\xi_{hh}(|{\bf x' + r}||m',m'')\\
\nonumber&+&\left.g_1^{sat}(m')g_2^{sat}(m'')\int d^3{\bf x'} d^3{\bf x''}\,  u(x'|m',c_1) u(x''|m'',c_2)\xi_{hh}(|{\bf x'-x'' + r}||m',m'')\right]
\end{eqnarray}
Again $i=1$ refers to nodes and $i=2$ to filaments;  $g^j(m)$ is the average number of galaxies that lie in a halo of mass $m$, the superindex $j$ denote the type of galaxy: central ($j\rightarrow cen$) or satellite ($j\rightarrow sat$) (see \S \ref{section_hod_chm} for details); $\bar n_{gal}$ is the mean number density of galaxies, i.e. the total number of galaxies divided by the total volume,
\begin{equation}
\bar n_{gal}=\sum_{i=1}^2{\cal X}_i\,\bar n_{gal\, i}
\end{equation}
and $n_{gal\, i}$ is the mean number density of galaxies inside haloes of type $i$, i.e. the total number of galaxies inside haloes of type $i$ divided by the volume these haloes occupy,
\begin{equation}
\bar n_{gal\, i}=\int dm\, n_i(m)\left[ g_i^{cen}(m)+g_i^{sat}(m)\right]\,.
\end{equation}
As before we can define an effective large scale galaxy bias as,
\begin{equation}
 b_{gal}=\sum_{i=1}^2{\cal X}_i \int dm \frac{n_i(m)}{\bar n_{gal}}\left(g_i^{cen}(m)+g_i^{sat}(m)\right)b(m)\,.
 \label{b_gal}
\end{equation}
In this case the effective large scale galaxy bias for nodes and filaments is given by,
\begin{equation}
 b_{gal}^i={\cal X}_i \int dm \frac{n_i(m)}{\bar n_{gal}}\left(g_i^{cen}(m)+g_i^{sat}(m)\right)b(m)\,.
\label{galaxy_bias}
\end{equation}
As before $b_{gal}=b_{gal}^{nod}+b_{gal}^{fil}$.
Note that for the galaxy bias, the difference between the node and the filament bias not only depends on the terms ${\cal X}_i$ and ${\cal Y}_i$ as in the dark matter bias (Eq. \ref{dm_bias}), but also on the HOD of each population. This means that even rescaling the mean number of galaxies $\bar n_{gal}$ to the mean number of galaxies of nodes or filaments, the biases will be different as we will see in the next section. 
The behaviour of these terms is illustrated in Fig. \ref{fig7}.  For simplicity $c_1=c_2=c(m)$ given by Eq. \ref{concentration} is adopted, along with the HOD for galaxies introduced by \cite{zehavi05}: $\log_{10}M_{min}=12.72$,  $\log_{10}M_1=14.08$ (the masses are in $M_\odot/h$), $\alpha=1.37$, $f_0^{cen}=0.71$, $f_0^{sat}=0.88$, $\sigma_M^{cen}=0.30$ and $\sigma_M^{sat}=1.70$  (see Appendix A \S \ref{section_hod_chm} for detailed definition of these HOD parameters) with maximum segregation: ${\cal S}=1$.
 This means that blue galaxies only populate filaments and red galaxies only populate nodes. The left panel shows the correlation functions of nodes (red line),  filaments (blue line), cross term (green line) and the total (black solid line).  In the right panel, the different lines correspond to the terms of Eq. \ref{ehm_gal_terms}: $1h1\eta$ (red line), $2h1\eta$ (blue line), $2h2\eta$ (green line) and the total contribution (black solid line). For comparison, the dark matter correlation function (black-dashed line) and the linear power spectrum (black-dotted line) are  also shown.

 In the left panel of Fig. \ref{fig7} we show the effect on the total correlation function (black solid line) of red galaxies (red line), blue galaxies (blue line) and the cross term (green line). We see that according to this HOD, red galaxies dominate the 1-halo term ($r<4$ Mpc/h), whereas all three terms contribute to the 2h term, being the cross term the most important.

\begin{figure}
\centering
\includegraphics[scale=0.49]{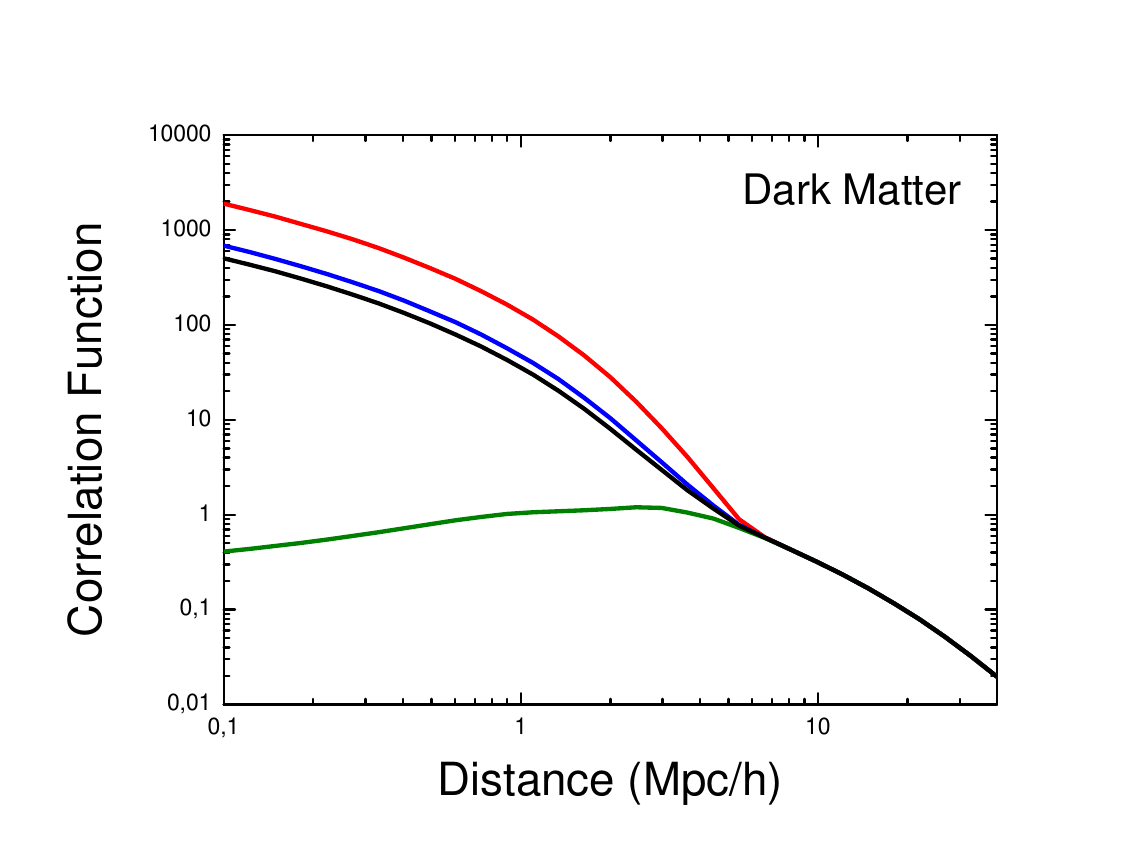}
\includegraphics[scale=0.49]{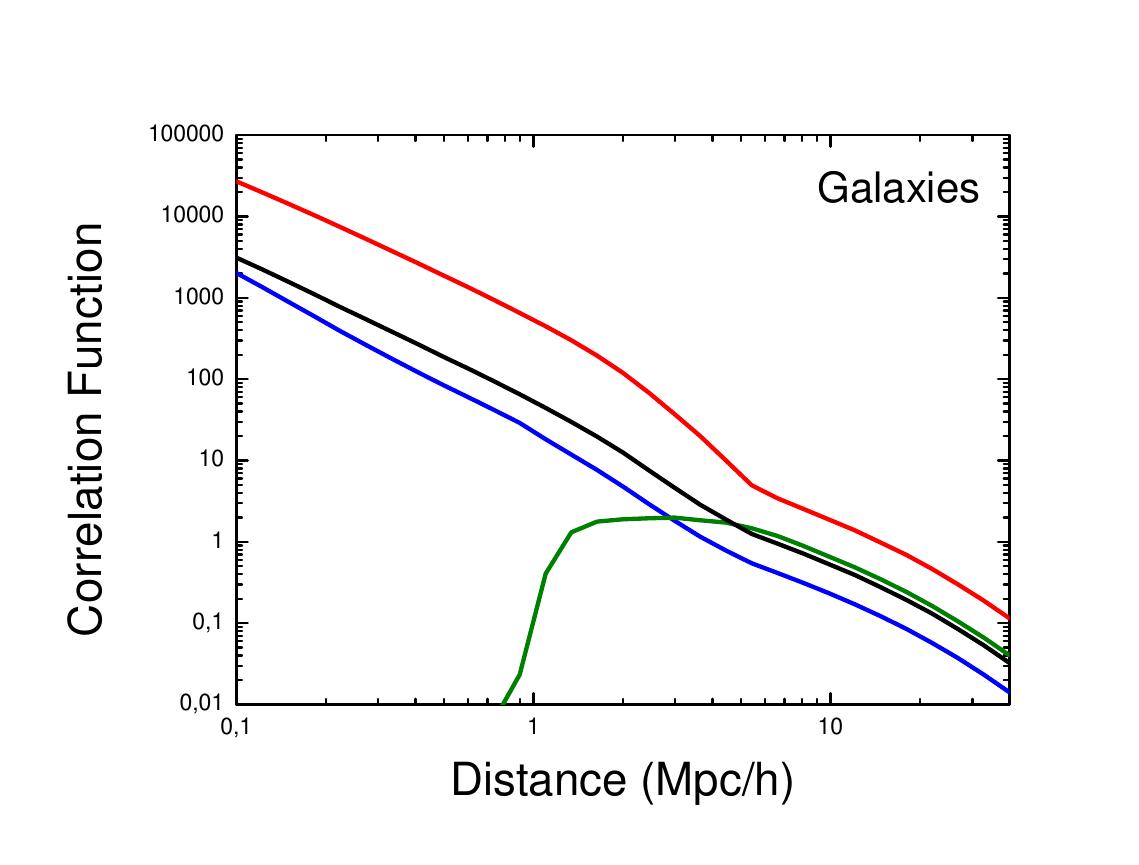}
\includegraphics[scale=0.49]{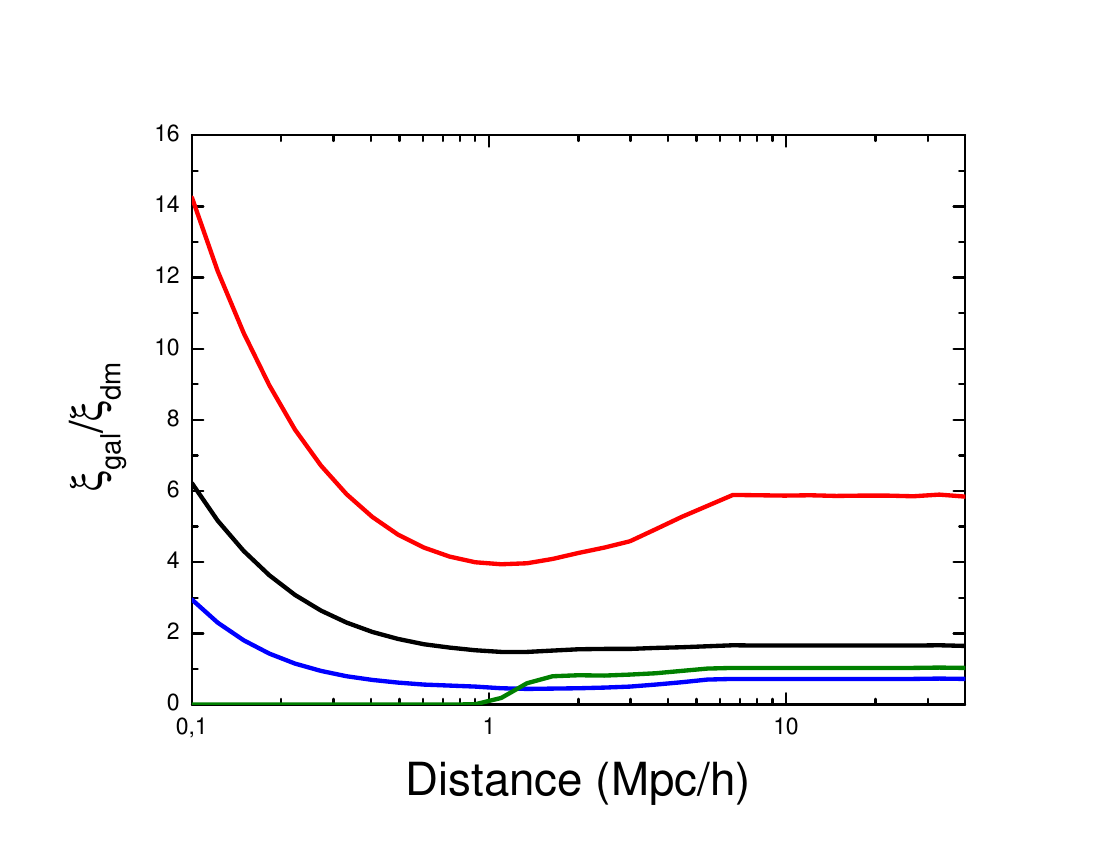}
\caption{Separate contribution for node- (red lines), filament-like (blue lines) haloes, the cross term (green line) for dark matter (left panel) and galaxies (central panel). The ratio between galaxies and dark matter correlation function is shown in right panel. Black lines correspond to the total contribution (node+filaments+cross).}
\label{plots_hods}
\end{figure}

We can also compute the `pure' node and filament galaxy terms, as we did in the dark matter case. These terms are shown in Fig. \ref{plots_hods} (central panel). As before, the shape is the same in both Figures but in the second one, the lines  normalisation is offset by $({\cal X}_i{\cal Y}_i)^{-2}$. In the case of the $2h2\eta$ term, the rescaling goes as $\xi^{2h2\eta}_{gal}\rightarrow\xi^{2h2\eta}_{gal}/(2{\cal X}_1{\cal Y}_1{\cal X}_2{\cal Y}_2)$ in order to obtain cross bias at large scales defined as $b^{cross}\equiv\sqrt{b^{nod}b^{fil}}$. This is due to the mean density definition as it is explained below.

In Fig. \ref{plots_hods} we show the 'pure' contribution of the node and filament terms for dark matter (left panel) and galaxies (central panel).  The correlation function for nodes/red galaxies is considerably higher than that for filaments/blue galaxies at all scales. This is just an effect of rescaling the mean density/number of galaxies from the total mass/galaxies over the total volume to the mass/galaxies in filaments or nodes over the total volume. This rescaling goes like ${\cal X}_i{\cal Y}_i$ for the mean density/number of galaxies and like (${\cal X}_i{\cal Y}_i)^{-2}$ for $\xi$. Since at small scales $\xi\sim{\cal X}_i{\cal Y}_i$, the ratio between the node and the filament 'pure' $\xi$ terms is just $({\cal X}_f{\cal Y}_f)/({\cal X}_n{\cal Y}_n)\simeq2.5$ and the pure correlation function for nodes is 2.5 times higher than that of filaments at small scales in the case of dark matter and a bit different in the case of galaxies due to the different HOD, as it can be seen in Fig. \ref{plots_hods} (left and central panel). However at large scales $\xi\sim({\cal X}_i{\cal Y}_i)^2$, which compensates for  the rescaling of the mean density/number of galaxies. Thus, in the case of dark matter  both filament and node pure correlation function are exactly the same since both effective large scale bias are the same after the rescaling (see Eq. \ref{dm_bias}).  However in the case of the galaxy correlation function, the effective large scale bias is considerably different for nodes and filaments due to they have different HOD (see Eq. \ref{galaxy_bias}), being higher for red galaxies than for blue ones.
In the right panel the ratio $\xi_{gal}/\xi_{dm}$ is shown for nodes (red line), for filaments (blue line), for the cross term (green line) and for the whole sample (black line). Both at small and large scales the ratio is higher for node-like haloes.

\section{Analysis}

In this section we perform a more exhaustive analysis to how HOD parameters, segregation and concentration affect to the dark matter and galaxy correlation function. We analyse how every parameter of the HOD affect the dark matter and galaxy correlation function while the other parameters  are kept fixed. We also analyse how the concentration parameter and the segregation affect to the both dark matter and galaxy correlation function while any other parameter is fixed. To gain insight we first  start with the standard halo model (i.e.  ${\cal S}=0$ and only one HOD prescription for al haloes) and then we move to our extended halo model with environmental dependence.

\subsection{Halo Model}

Here we adopt Eqs. \ref{HOD_cen} and \ref{HOD_sat} as a parametrisation of the HOD and we analyse how the different  parameters ($M_{min}$, $M_1$ and $\alpha$) affect $\xi_{gal}$. Here $M_{min}$ sets the minimum mass for a halo to have galaxies, $M_1$ is the mass of a halo that on average hosts one satellite galaxy and $\alpha$ is the power-law slope of the satellite mean occupation function (see Appendix A \S \ref{section_hod_chm} for more details).
We set the parameters to  fiducial values which are very close to the ones proposed by \cite{kravtsov04}: $M_{min}=10^{11} M_\odot$, $M_1=22\times10^{11} M_\odot$ and $\alpha=1.00$. In the following section we analyse how changing each one of these parameters affects  the galaxy correlation function and the ratio $\xi_{gal}/\xi_{dm}$. Since filament galaxies are the most abundant, the effects we find in this case would be qualitatively very similar to the effect of changing the HOD of filament population, keeping the nodes HOD  fixed at the fiducial model.

\subsubsection{Central Galaxies}
\begin{figure}
 \centering
\includegraphics[scale=0.7]{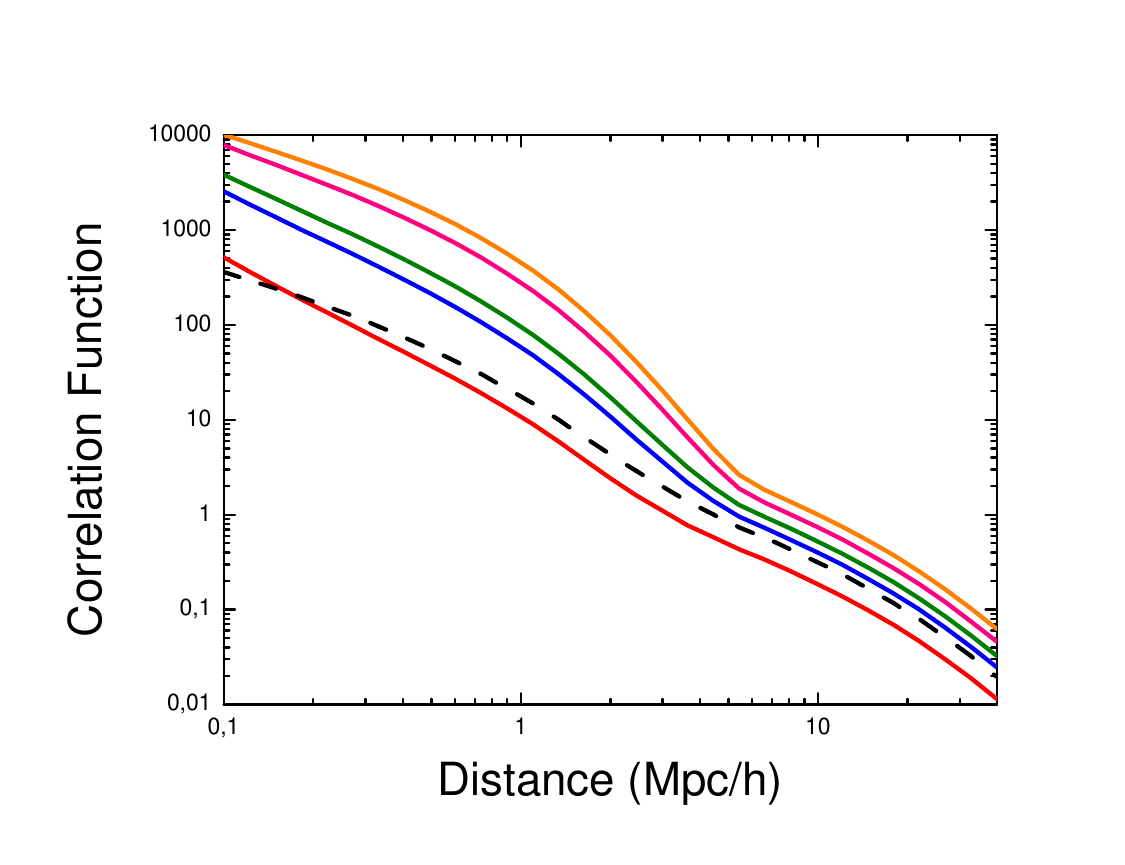}
\includegraphics[scale=0.7]{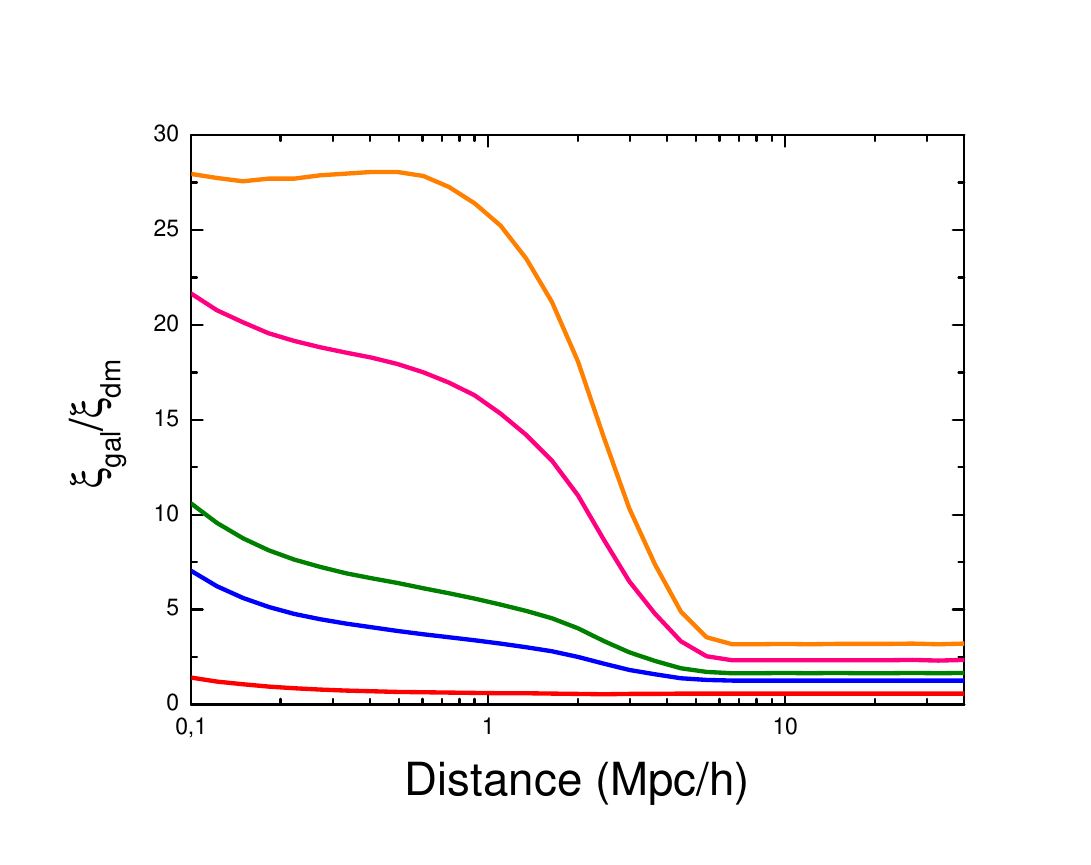}
\caption{Effect of changing $M_{min}$ in the HOD model. In the right panel  the galaxy correlation function $\xi_{gal}$ is shown, whereas in the left panel the ratio $\xi_{gal}/\xi_{dm}$ is shown. The different colours correspond to  different $M_{min}$ values: $10^{11}$ $M_\odot$ (red line), $5\times10^{11}$ $M_\odot$ (blue line), $10^{12}$ $M_\odot$ (green line), $5\times10^{12}$ $M_\odot$ (pink line) and $10^{13}$  $M_\odot$ (orange line). In the left panel, black-dashed line corresponds to  the dark matter correlation function $\xi_{dm}$.}
\label{mmin_plot}
\end{figure}

We start by varying $M_{min}$, the minimum mass of a halo to host at least one galaxy, keeping all the other HOD parameters fixed at their fiducial values. 
 In Fig. \ref{mmin_plot}  the different solid lines correspond to  different values of $M_{min}$: $10^{11}$ (red line),  $5\times10^{11}$ (blue line), $10^{12}$ (green line),  $5\times10^{12}$ (pink line) and $10^{13} M_\odot$ (orange line). The left panel is the galaxy correlation function and the right panel is $\xi_{gal}/\xi_{dm}$. For reference,  in the left panel, the dark matter correlation function $\xi_{dm}$ is  also plotted (black-dashed line).

We can see that at large scales the galaxy correlation function is just a boost of the dark matter correlation function, whereas at small scales the shape is modified. The enhancement of the amplitude of the correlation function at large scales can be explained just as a redistribution of galaxies: removing galaxies from low mass haloes ($M<M_{min}$) causes a large-scale bias increase because the function $b(m)$ (Eq. \ref{bias}) increases with the mass. Therefore at these large scales we expect that only the amplitude of the galaxy correlation function changes with $M_{min}$ (and not the shape that is given by $\xi_{lin}$). On the other hand at small scales the 1-halo term of the galaxy correlation function becomes important; this term does not depend explicitly on $\xi_{dm}$, therefore the shape of the correlation function does not have to be the same. The behaviour of Fig. \ref{mmin_plot} is qualitatively very similar to the findings of \cite{berlind_weinberg}.

\subsubsection{Satellite Galaxies}

Here we keep fix $M_{min}$ and $\alpha$ to the  fiducial values $10^{11} M_\odot$ and $1.00$ respectively and we vary $M_1$ from $22\times10^{10} M_\odot$ (red line) to $10^{13} M_\odot$ (orange line). Recall that this is the minimum  mass  for a halo  to  host at least one  satellite galaxy.

\begin{figure}
 \centering
\includegraphics[scale=0.7]{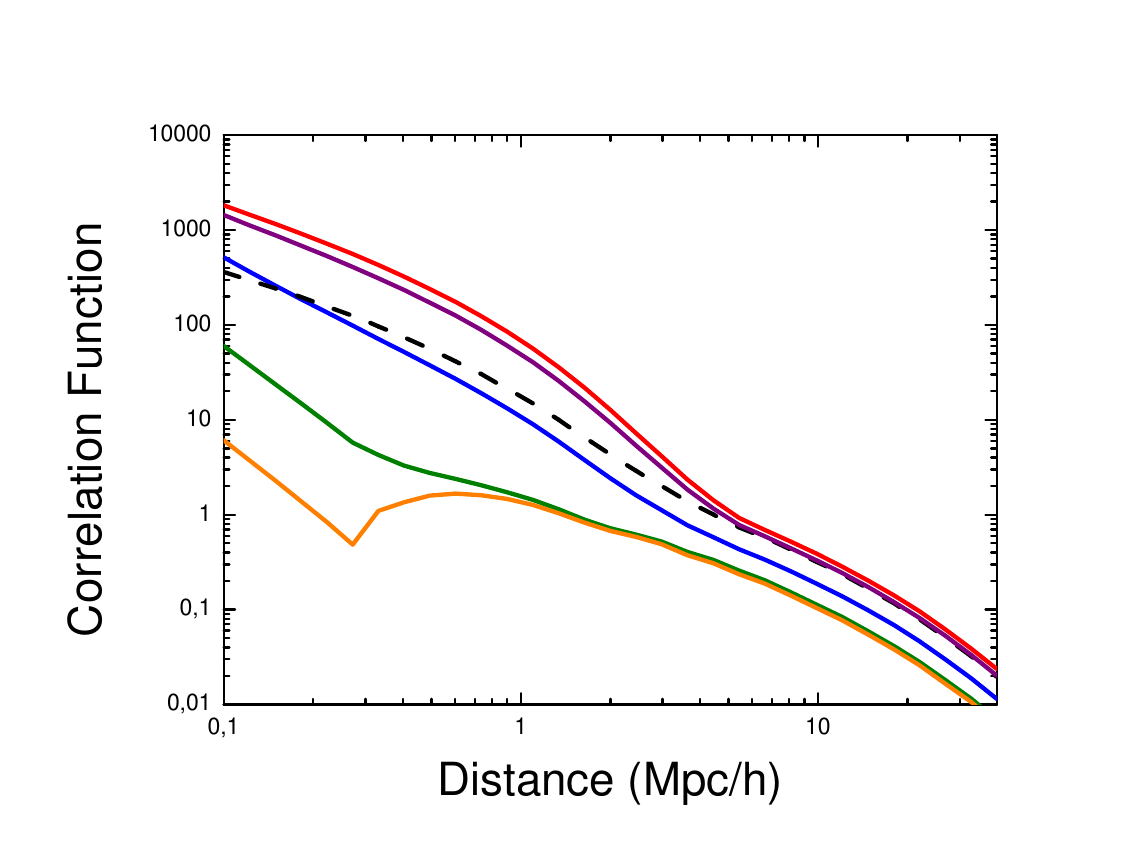}
\includegraphics[scale=0.7]{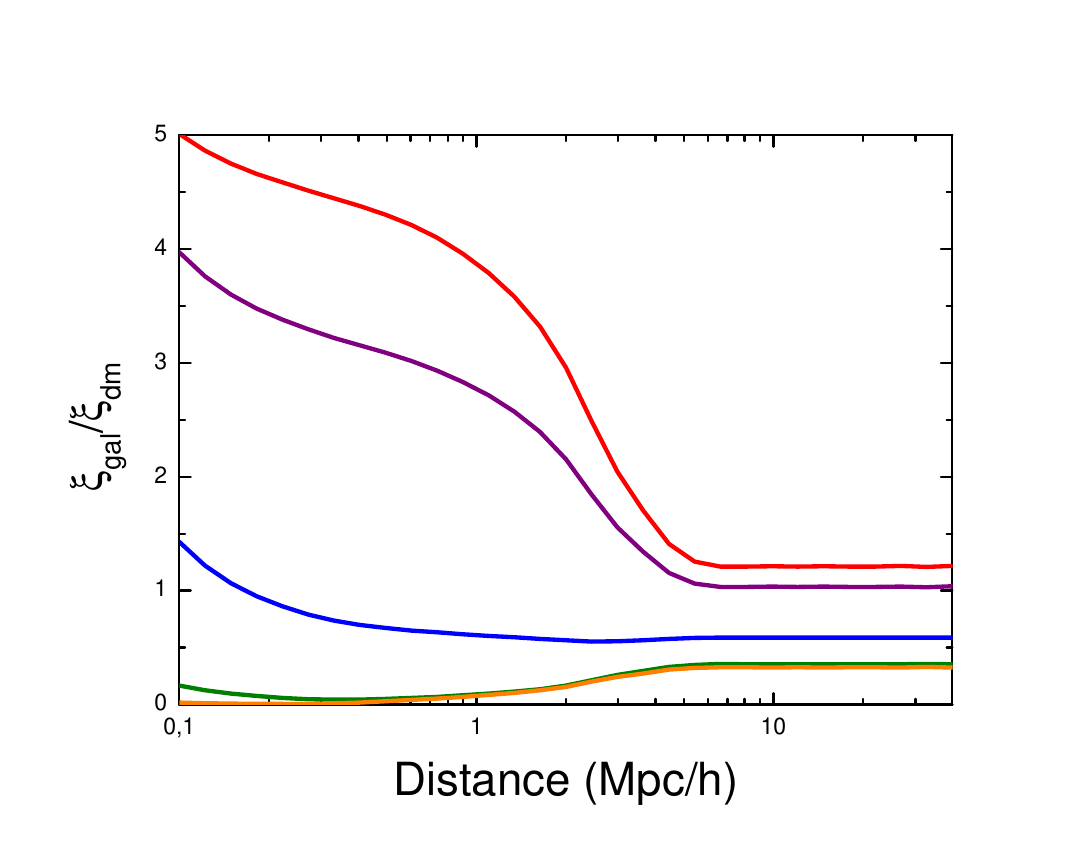}
\caption{Effects of changing $M_1$ in the halo model. As in Fig. \ref{mmin_plot} the left panel is $\xi_{gal}$ and the right panel $\xi_{gal}/\xi_{dm}$. Different colours are different values of $M_1$: $22\times10^{10}$ $M_\odot$ (red line), $44\times10^{10}$ $M_\odot$ (purple line), $22\times10^{11}$ $M_\odot$ (blue line), $22\times10^{12}$ $M_\odot$ (green line) and $22\times10^{13}$ $M_\odot$ (orange line). In left plot $\xi_{dm}$ is also represented (black dashed line).}
\label{m1_plot}
\end{figure}

In Fig. \ref{m1_plot} we show the effect of varying this parameter. The left panel shows the galaxy correlation function and the right panel $\xi_{gal}/\xi_{dm}$ for different values of $M_1$ (see caption for details). In the left panel, $\xi_{dm}$ is  also plotted (black-dashed line). At sub-halo scales we observe that  high values of  $M_1$  suppress the  correlation function. This effect is expected: the higher $M_1$, the less satellites galaxies per halo and therefore the  1-halo term is suppressed.  We also observe that for very high values of $M_1$ ($>22\times10^{12} M_\odot$) at very small scales $\xi_{gal}$  shows an inflection point around $0.3\,\mbox{Mpc}/h$.
This is due to the interaction between the central galaxies of different haloes. Since for these values of $M_1$ there is (almost) no satellite galaxy per halo, we can observe the interaction between central galaxies of the smallest haloes (those of a mass $\sim M_{min}$), which are the only ones that at that distances can contribute to the 2-halo term on those scales. Such feature is similar to the one of the 2-halo term of Fig. \ref{fig7}, just with the difference that in that case $M_{min}\sim10^{12.72} M_\odot$, whereas now $M_{min}=10^{11} M_\odot$ and therefore such feature happens at shorter distances.

At large scales we also see that the correlation decreases as we increase $M_1$. This is also expected: according to Eq. \ref{b_gal} , low mass haloes are weighted more in contributing to the large scale bias, as one decreases $M_1$ the large scale correlation function also decreases. In other words, for high values of $M_1$ all haloes count the same (1 central galaxy per halo) independently of their mass.

Now we consider the variation of $\alpha$ from 0.8 to 1.4, keeping  $M_{min}=10^{11}$ $M_\odot$ and $M_1=22\times10^{11}$ $M_\odot$. This parameter indicates how the number of satellite galaxies increase with the mass of the halo. A steeper  slope $\alpha$ indicates more extra satellites per halo mass increment.

\begin{figure}
 \centering
\includegraphics[scale=0.7]{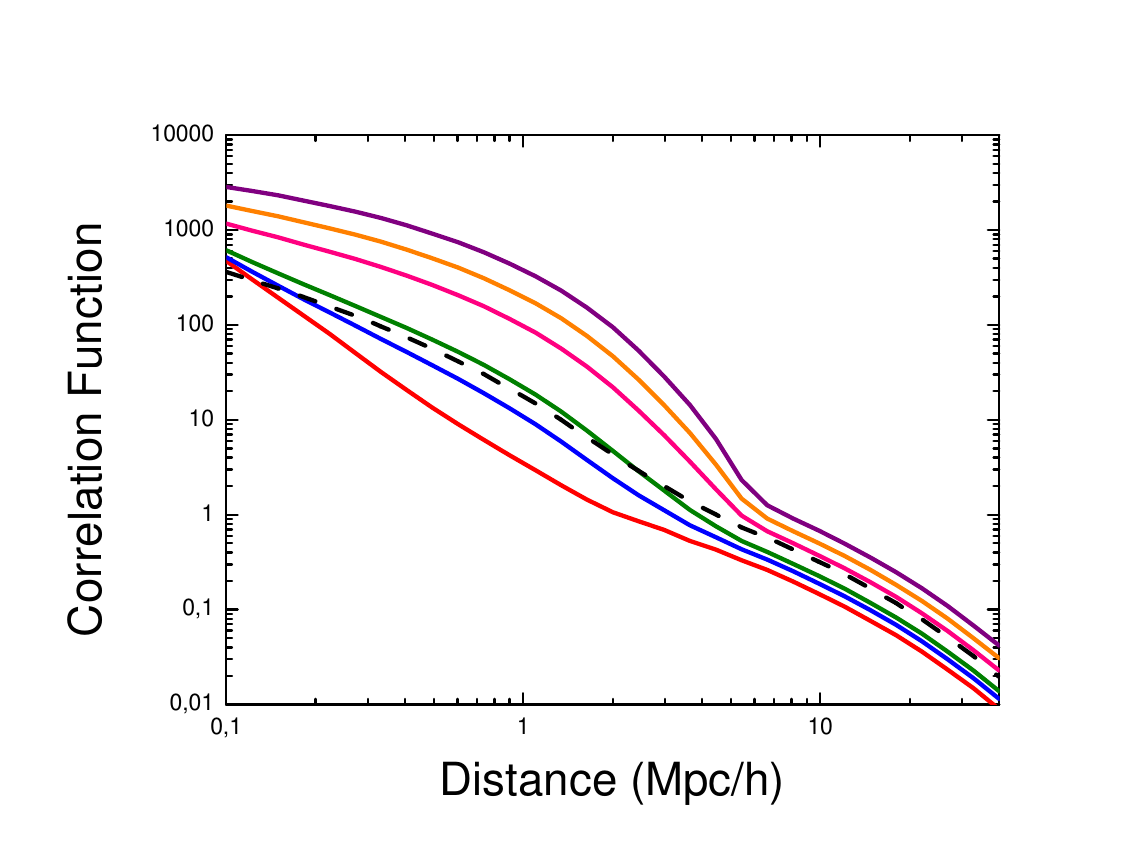}
\includegraphics[scale=0.7]{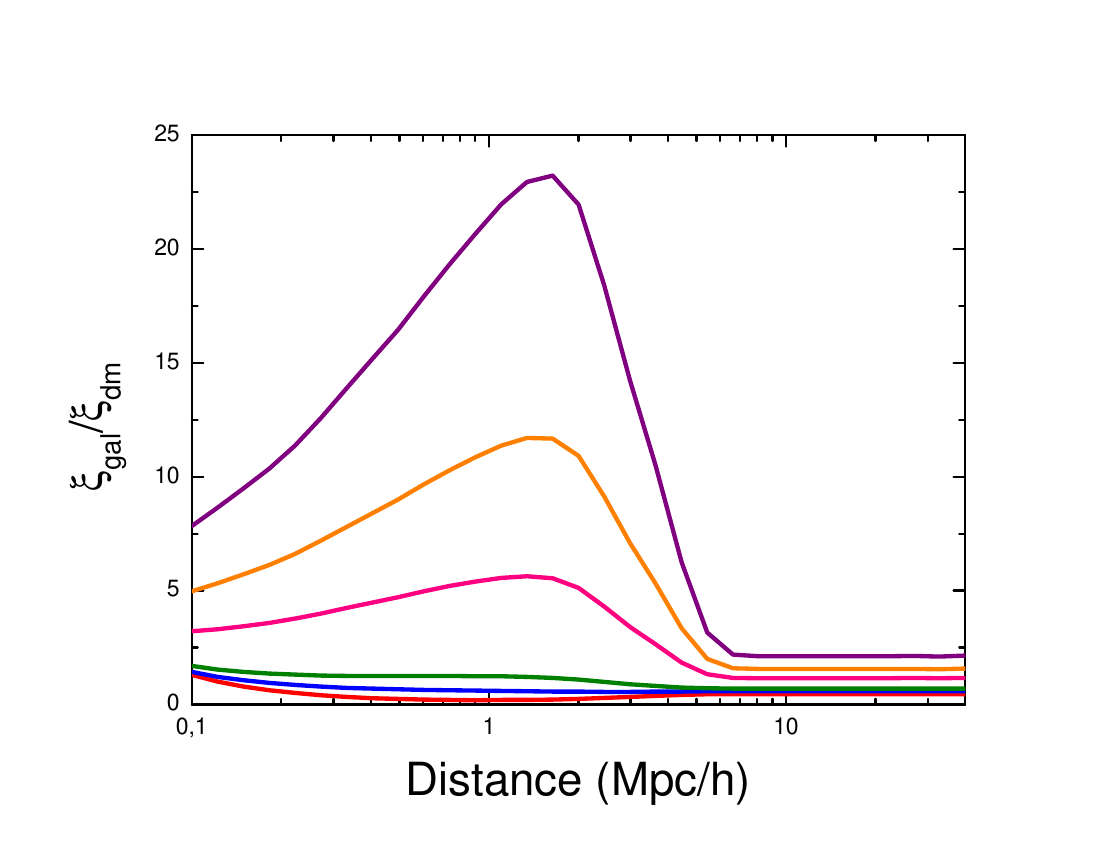}
\caption{Effects of changing $\alpha$ in the halo model. As in Fig. \ref{mmin_plot} left panel is $\xi_{gal}$ and the right panel $\xi_{gal}/\xi_{dm}$. Different colours are different values of $\alpha$: 0.80 (red line), 1.00 (blue line), 1.10 (green line), 1.30 (pink line), 1.40 (orange line) and 1.50 (purple line). In the left plot $\xi_{dm}$ is the black-dashed plot.}
\label{alpha_plot}
\end{figure}

In Fig. \ref{alpha_plot} we show the effect of changing $\alpha$ in the galaxy correlation function (left panel) and in $\xi_{gal}/\xi_{dm}$ (right panel). The different colours  indicate different values for $\alpha$ (see caption for details) and the black-dashed line is the dark matter correlation function.

As expected, when   $\alpha$ increases, the galaxy correlation function also increases, especially at sub-halo scales. At large scales the effect of increasing $\alpha$ is the same as increasing $M_{min}$: it boosts the dark matter correlation function without changing its shape. This is because increasing $\alpha$ we weight more the more massive haloes and therefore the bias increases.
At small scales there is also an enhancement of the correlation function for the same reason, but in this case the shape of $\xi_{gal}$ is qualitatively different from the shape of $\xi_{dm}$, because as in the case of changing $M_{min}$, there is not direct relation between the 1-halo terms of $\xi_{gal}$ and $\xi_{dm}$. This behaviour was also noted by \cite{berlind_weinberg}.

\subsection{Extended Halo Model}
In this section we use again the HOD parametrised by Eq. \ref{segregation} using as $g^{cen}(m)$ and $g^{sat}(m)$ for filament- and node-like haloes Eqs. \ref{HOD_cen} and \ref{HOD_sat} but with different parameters. In particular we keep fixed the HOD parameters for filament haloes to the fiducial values: $M_{min}=10^{11} M_\odot$, $M_1=22\times10^{11} M_\odot$, $\alpha=1$ and we allow to vary $M_{min}$, $M_1$ and $\alpha$ only for red galaxies. We assume a concentration $c(m)$ given by Eq. \ref{concentration} and  begin with a maximum segregation index ${\cal S}=1$ (this will later be relaxed). Since the filaments are the most common structure, changing their HOD would give a similar effect as  changing  the HOD of the whole population, that we studied in the previous  section. Therefore, in this section we focus on how changing the HOD of the less abundant population (in this case the nodes) can affect to the total correlation function.

\subsubsection{Central Galaxies in nodes}
We set $M_1=22\times10^{11} M_\odot$ and $\alpha=1.0$ for node-like satellites and we vary $M_{min}$ from $10^{11}$ $M_\odot$ to $10^{13} M_\odot$.

\begin{figure}
 \centering
\includegraphics[scale=0.7]{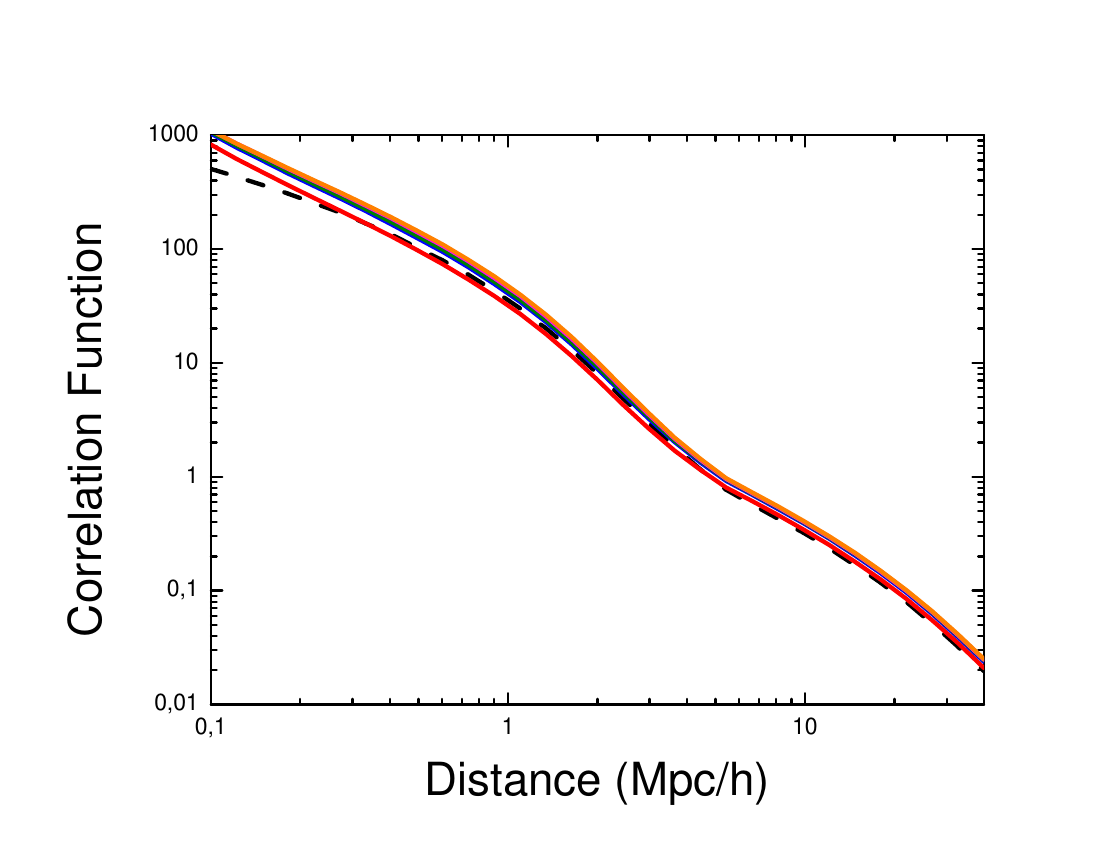}
\includegraphics[scale=0.7]{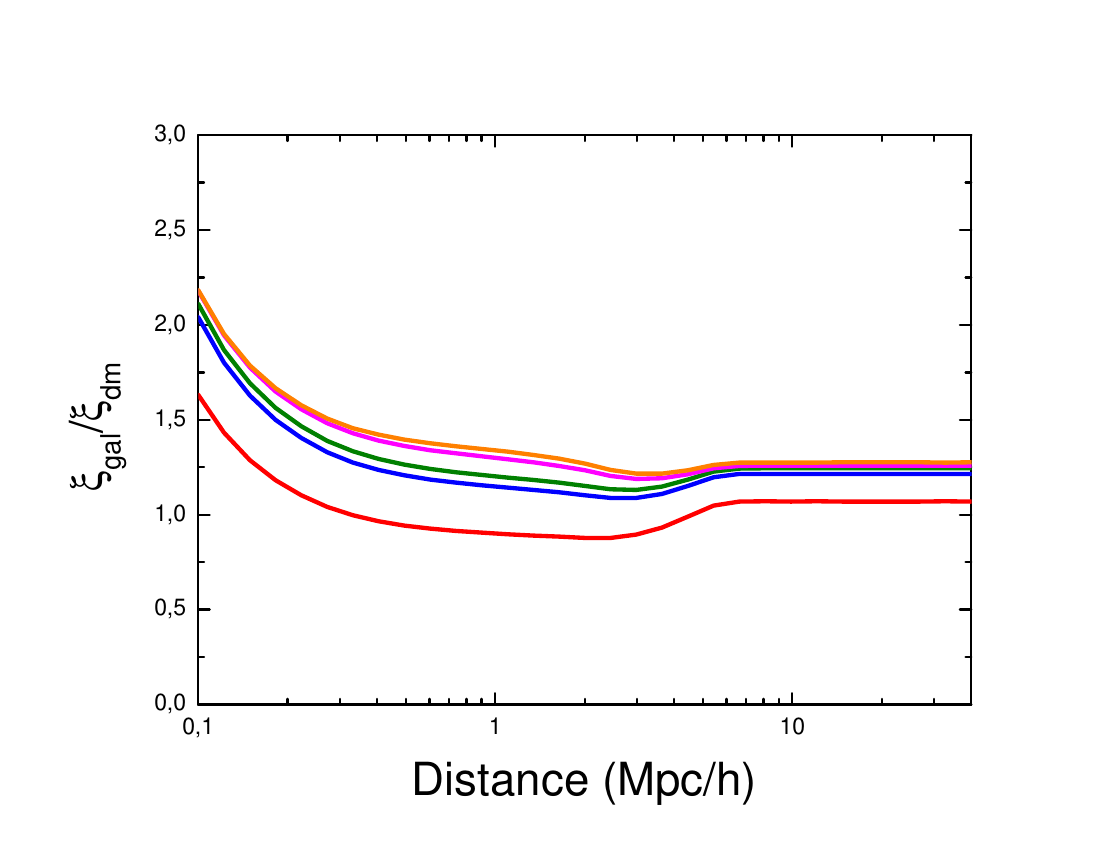}
\caption{Effect of changing $M_{min}$ for node-like haloes in the extended halo model. Left panel is $\xi_{gal}$; right panel is $\xi_{gal}/\xi_{dm}$. The different colours are different values for $M_{min}$: $10^{11}$ $M_\odot$ (red line), $5\times10^{11}$ $M_\odot$ (blue line), $10^{12}$ $M_\odot$ (green line), $5\times10^{12}$ $M_\odot$ (pink line) and $10^{13}$ $M_\odot$ (orange line). In the left panel  $\xi_{dm}$ is also plotted (black-dashed line). }
\label{mmin_plot_ehm}
\end{figure}

In Fig. \ref{mmin_plot_ehm} we show the effect of changing $M_{min}$ for node-like galaxies. The different colour lines are $\xi_{gal}$ for different values of $M_{min}$ for node-like galaxies (see caption for details) and the black-dashed line is $\xi_{dm}$.
In the left panel we show the total correlation function. We see that the total galaxy correlation function is not very sensitive to this parameter. We observe only a small to moderate change, much less that that  observed when we changed $M_{min}$ of the total population of haloes. This can be due to the fact that if we increase $M_{min}$ less node-like haloes are populated, so there are less red galaxies and the total correlation function is dominated by the filament-like galaxies. In the right plot we show the corresponding effect on the $\xi_{gal}/\xi_{dm}$.

\subsubsection{Satellite Galaxies in nodes}
Here we keep  $M_{min}=10^{11} M_\odot$ and $\alpha=1.0$ fixed and we allow $M_1$ to vary. 
\begin{figure}
 \centering
\includegraphics[scale=0.7]{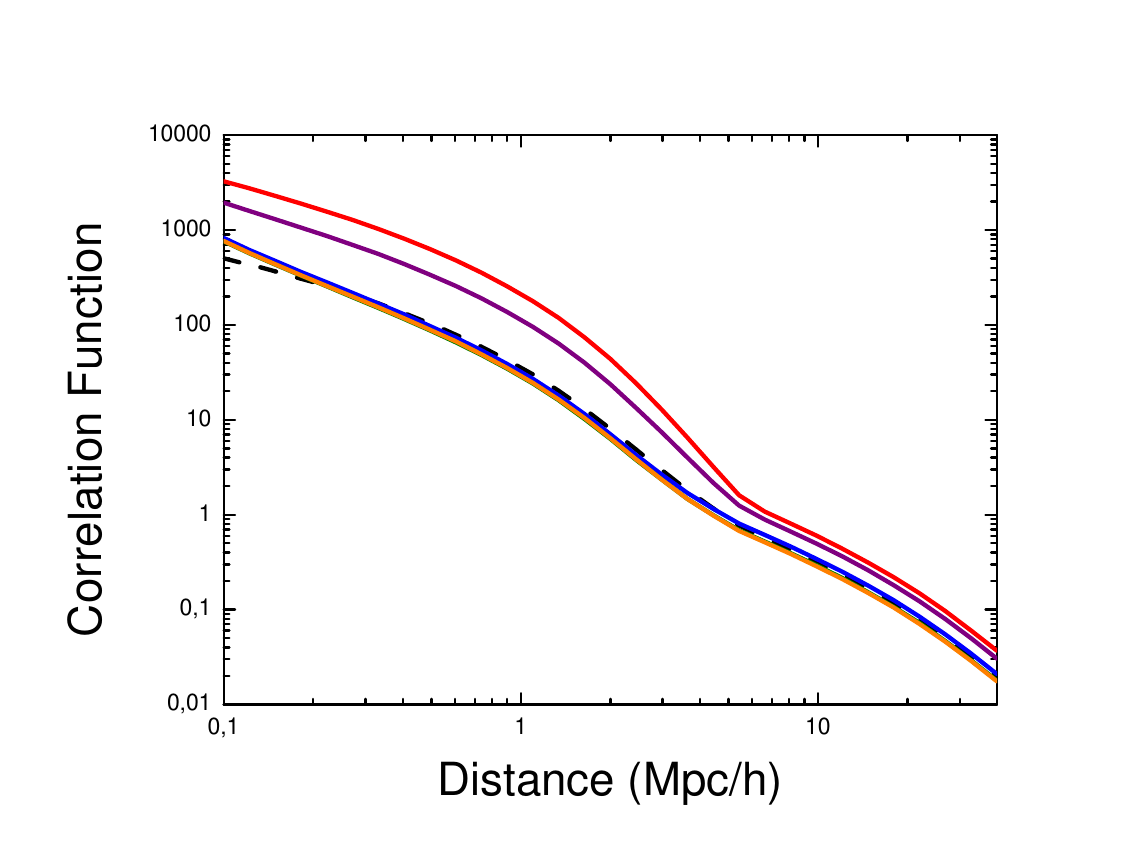}
\includegraphics[scale=0.7]{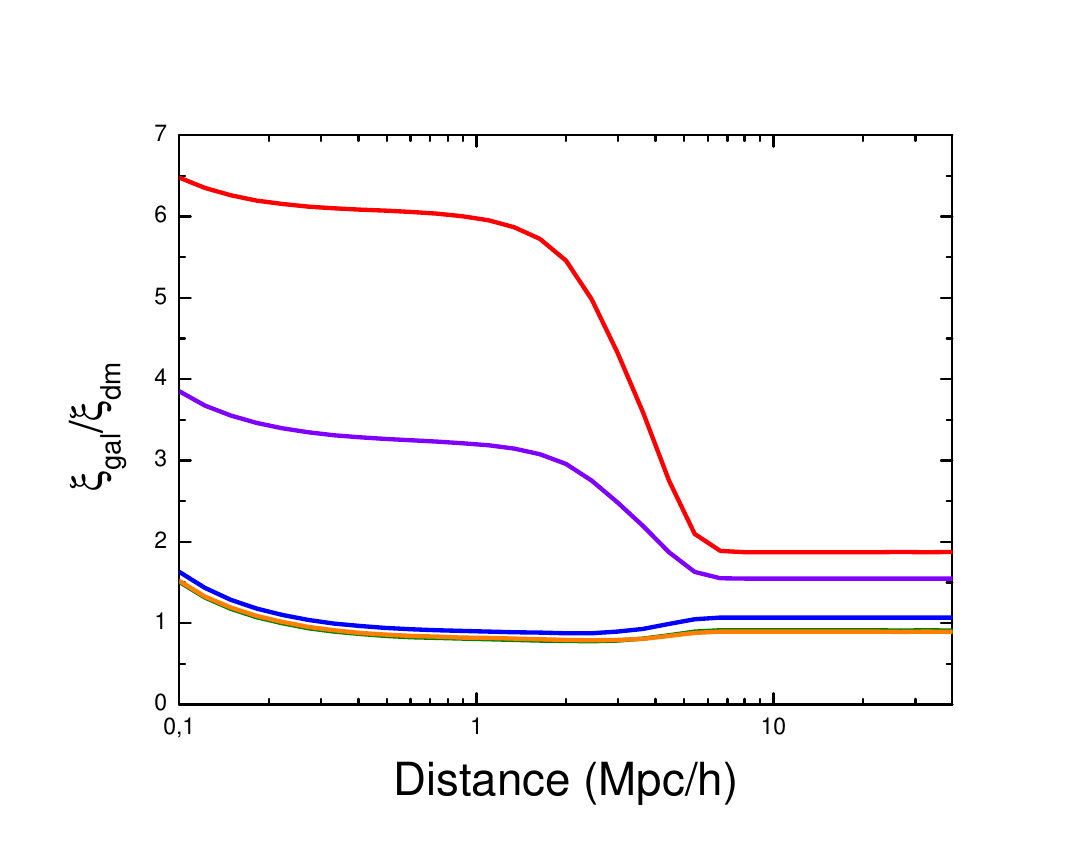}
\caption{Effect of changing $M_1$ in node-like haloes. As in Fig. \ref{mmin_plot_ehm} in left panel $\xi_{gal}$ is shown and in the right panel $\xi_{gal}/\xi_{dm}$ is plotted. The different colours correspond to  different values for $M_1$: $22\times10^{10}$ $M_\odot$ (red line), $44\times10^{10}$ $M_\odot$ (purple line), $22\times10^{11}$ $M_\odot$ (blue line), $22\times10^{12}$ $M_\odot$ (green line) and $22\times10^{13}$ $M_\odot$ (orange line). In the left panel, the black-dashed line represents $\xi_{dm}$.}
\label{m1_plot_ehm}
\end{figure}
In Fig. \ref{m1_plot_ehm} we show the effect of changing $M_1$ in the galaxy correlation function (left panel) and on $\xi_{gal}/\xi_{dm}$ (right panel). The different colour lines represent different values of $M_1$. As before, the black-dashed line is the dark matter correlation function. Recall that the effect of increasing $M_1$ is to reduce the number of  satellite galaxies for low-mass haloes in the node-like regions. We see that $\xi_{gal}$ is only sensitive to the change of $M_1$ in the range from $22\times10^{10} M_\odot$ to $22\times10^{11} M_\odot$. For values larger than $22\times10^{11} M_\odot$ the correlation function saturates. In particular we see that $\xi_{gal}$ is especially sensitive to $M_1$ at small scales where the 1-halo term dominates. This means that the number of satellites in the low-mass and node-like haloes plays an important role in the total correlation function.

\begin{figure}
 \centering
\includegraphics[scale=0.7]{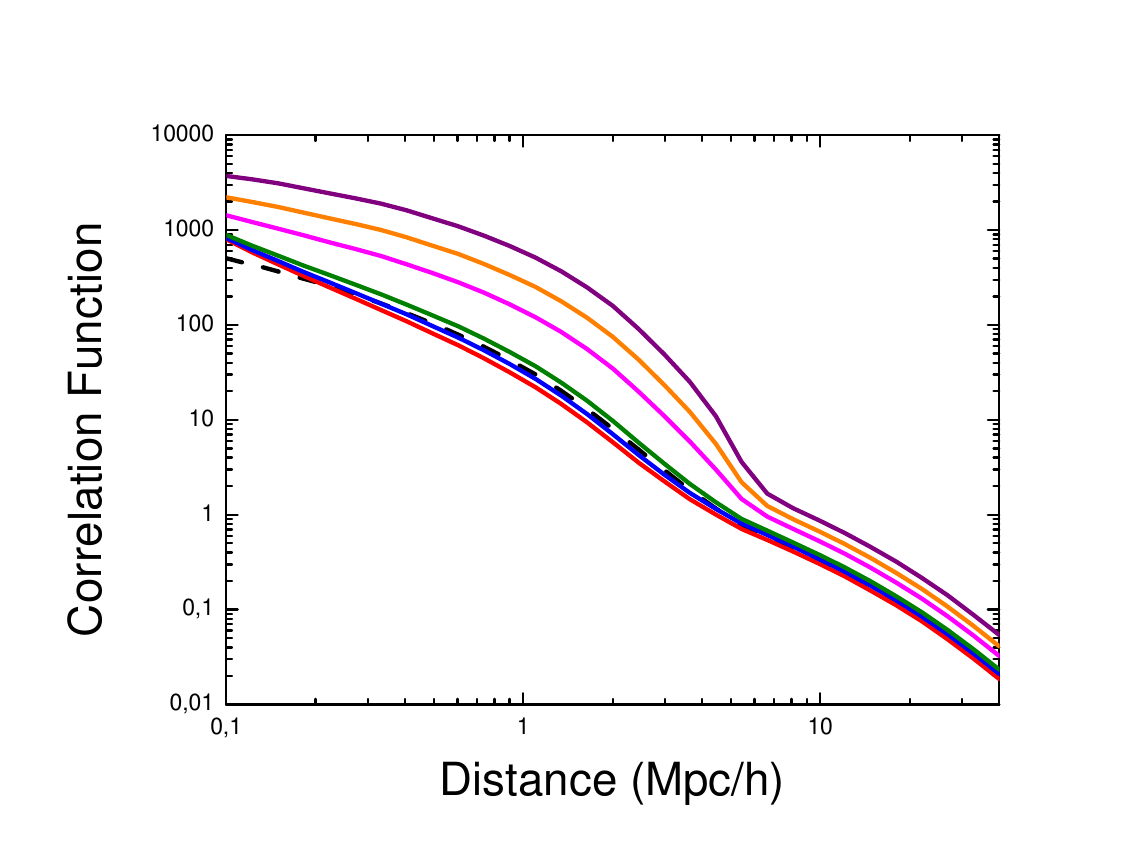}
\includegraphics[scale=0.7]{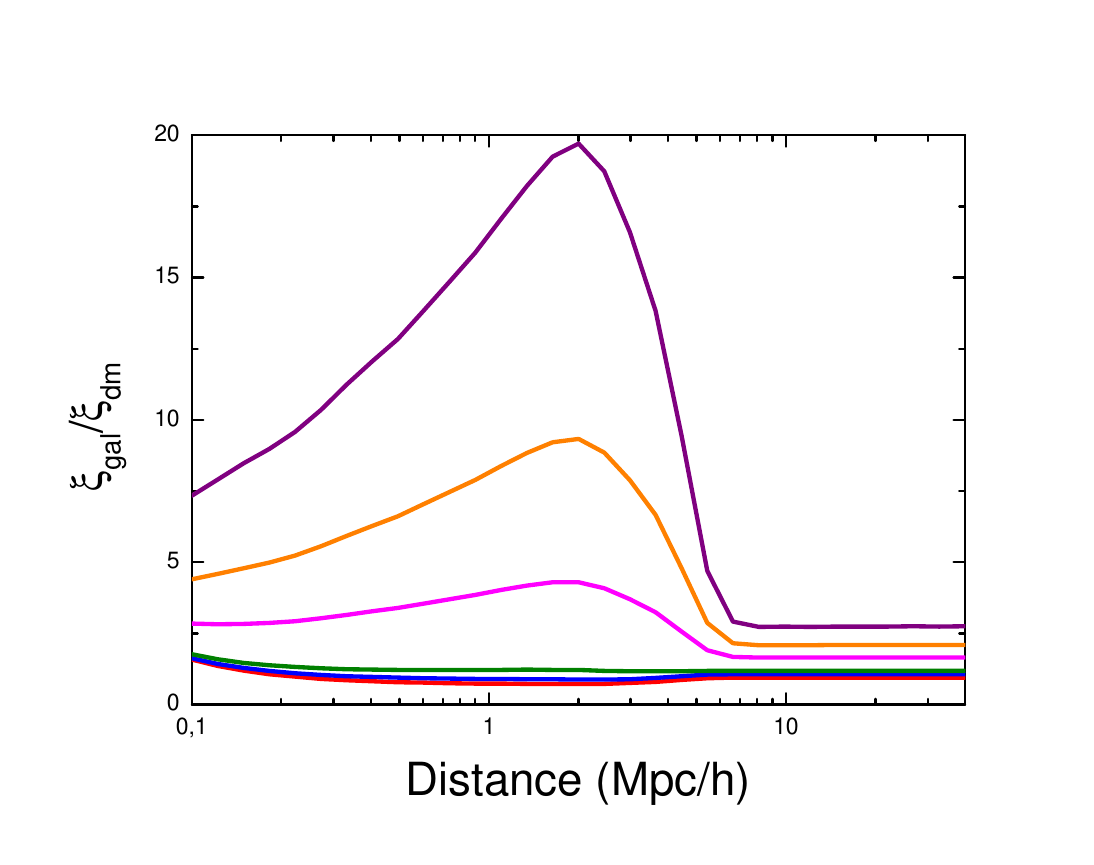}
\caption{Effect of changing $\alpha$ on nodes-like haloes in our halo model. As in Fig. \ref{mmin_plot_ehm} in left panel $\xi_{gal}$ is shown whereas in the right panel $\xi_{gal}/\xi_{dm}$ is plotted. The different colours show different values of $\alpha$: 0.80 (red line), 1.00 (blue line), 1.10 (green line), 1.30 (pink line), 1.40 (orange line) and 1.50 (purple line).  In the left plot, the black-dashed line represents $\xi_{dm}$.}
\label{alpha_plot_ehm}
\end{figure}

In Fig. \ref{alpha_plot_ehm} we show the effect of changing $\alpha$ in node-like galaxies keeping fixed $M_{min}$ and $M_1$ to $10^{11} M_\odot$ and $22\times10^{11} M_\odot$ respectively. We explore the regime from $\alpha=0.80$ (red line) to $\alpha=1.50$ (purple line) for the $\xi_{gal}$ (left panel) and for $\xi_{gal}/\xi_{dm}$ (right panel). As before, $\xi_{dm}$ is the black-dashed line. We observe the same effect observed in Fig. \ref{alpha_plot}. As in the $M_1$ case, $\xi_{gal}$ is especially sensitive to $\alpha$ at small distances where the 1-halo term dominates. This may indicate that for these values of ${\cal X}$ and ${\cal Y}$ the satellite galaxies in the node-like haloes have an important role at small scales on the total galaxy correlation function in spite of not being the dominant population.

\subsection{Concentration}
\begin{figure}
 \centering
\includegraphics[scale=0.49]{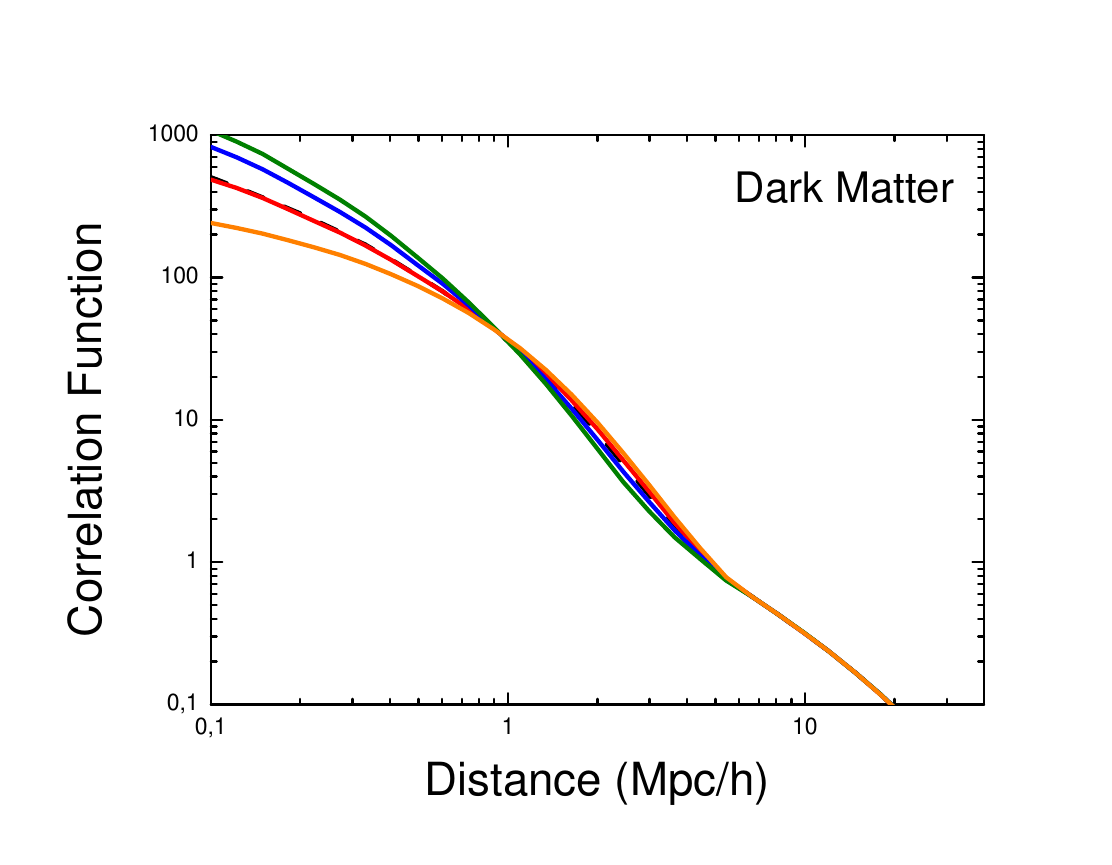}
\includegraphics[scale=0.49]{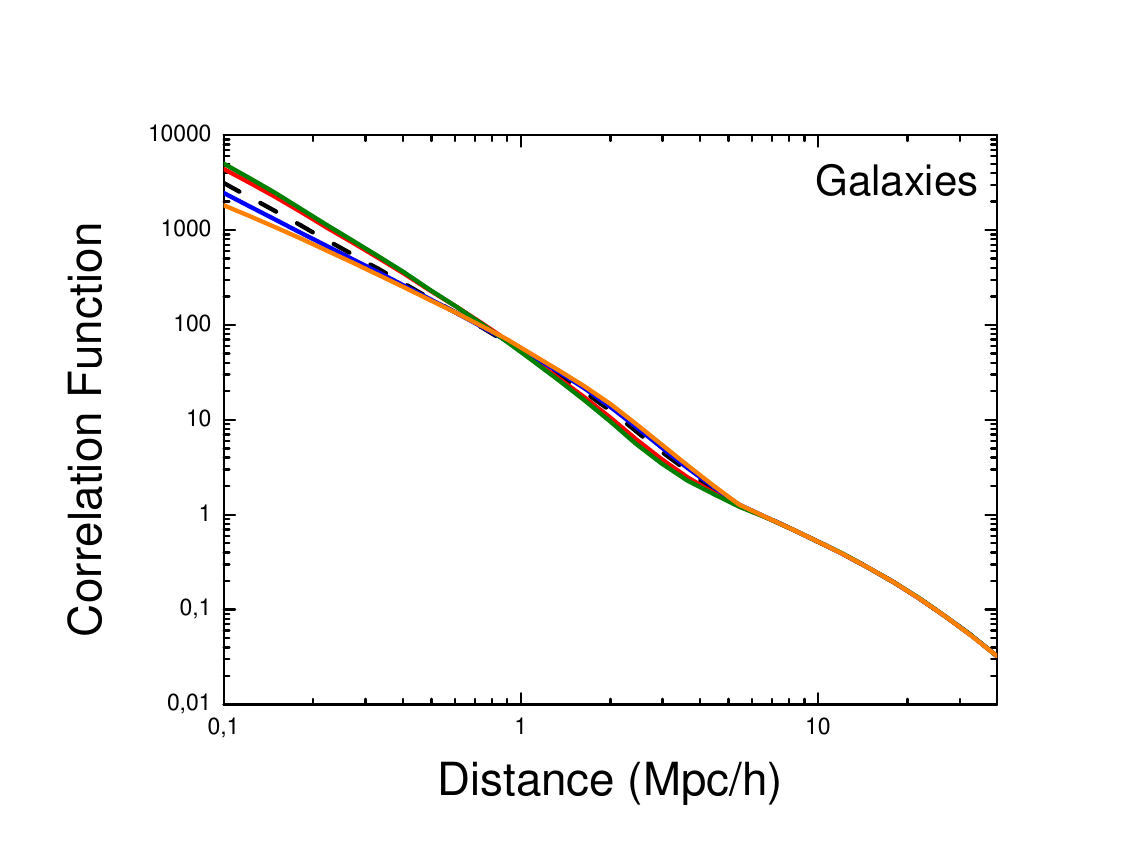}
\includegraphics[scale=0.49]{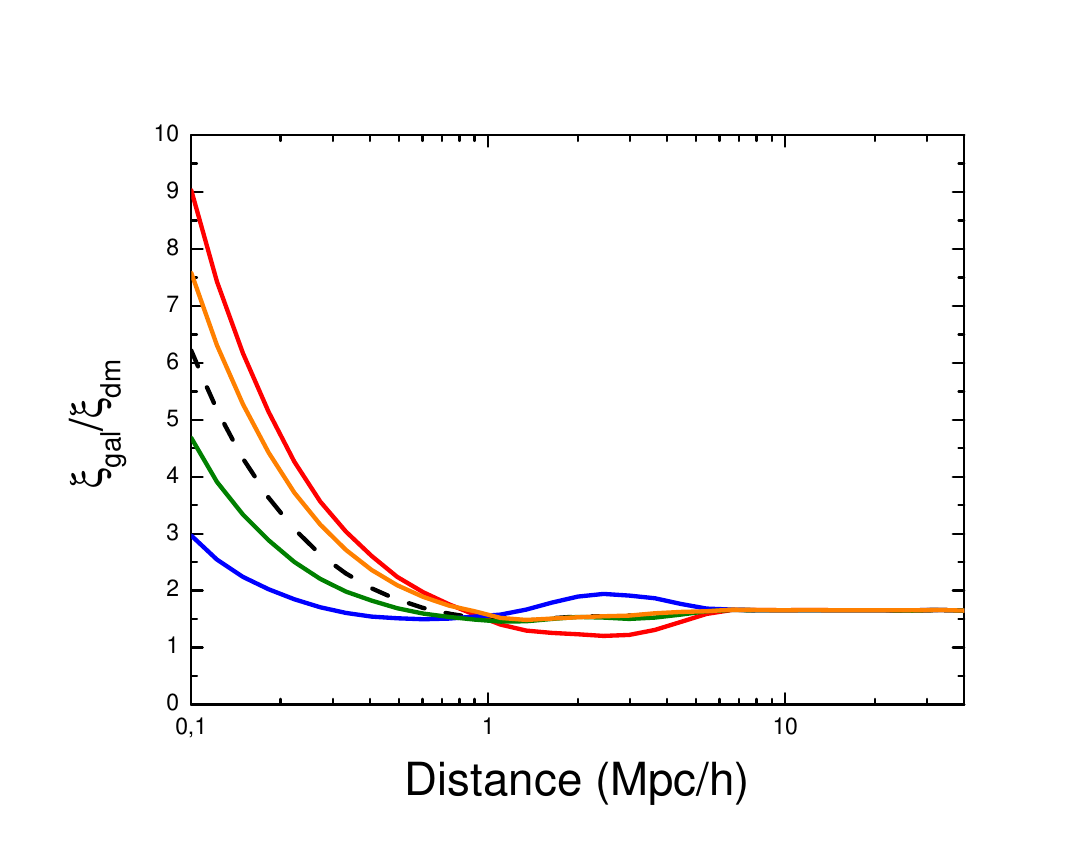}
\caption{Correlation function for different values of the concentration parameters: $c_{nod}=10$ \& $c_{fil}=2$ (red line),  $c_{nod}=2$ \& $c_{fil}=10$ (blue line),  $c_{nod}=10$ \& $c_{fil}=10$ (green line) and  $c_{nod}=2$ \& $c_{fil}=2$ (orange line). The black-dashed line represents the case for $c(m)$ according to Eq. \ref{concentration}. In the left panel, the dark matter correlation function is plotted, in the central panel the galaxy correlation function and in the right plot  $\xi_{gal}/\xi_{dm}$.}
\label{concentration_ehm}
\end{figure}

In Fig. \ref{concentration_ehm} we show the effect of changing the concentration parameter for both dark matter (left panel) and galaxy correlation function (central panel). Also in the right panel we show the effect on $\xi_{gal}/\xi_{dm}$. The different lines are for different concentration values of nodes and filaments: $c_{nod}=10$ \& $c_{fil}=2$ (red line),  $c_{nod}=2$ \& $c_{fil}=10$ (blue line),  $c_{nod}=10$ \& $c_{fil}=10$ (green line) and  $c_{nod}=2$ \& $c_{fil}=2$ (orange line). For comparison we also have plotted $c(m)$ according to Eq. \ref{concentration} (black dashed line). Here we have fixed the HOD to the one of \cite{zehavi05}:  $\log_{10}M_{min}=12.72$,  $\log_{10}M_1=14.08$ (the masses are in $M_\odot/h$), $\alpha=1.37$, $f_0^{cen}=0.71$, $f_0^{sat}=0.88$, $\sigma_M^{cen}=0.30$ and $\sigma_M^{sat}=1.70$  (see Appendix A \S \ref{section_hod_chm} for detailed definition of these HOD parameters) and  adopted the maximum value for the segregation index, ${\cal S}=1$. Note that significant changes appear at small scales ($r<1$ Mpc/h). Also these changes are only  important for the dark matter correlation function. On the other hand, for galaxies are almost negligible.

In general the effect of changing the concentration can be summarised up as a change in the slope of the the small scale correlation function (especially the dark matter one), with a more concentrated profile (green line) corresponding to a steeper correlation function.

\subsection{Segregation}
\begin{figure}
 \centering
\includegraphics[scale=0.7]{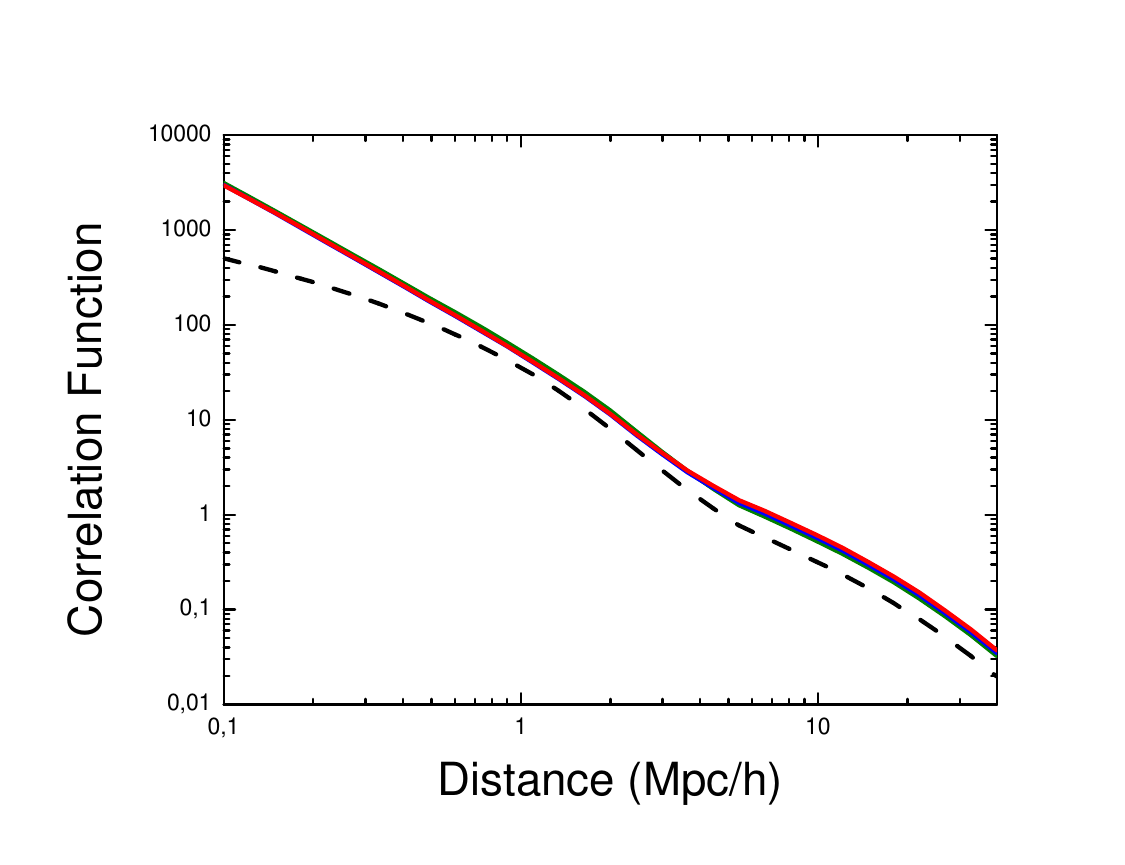}
\includegraphics[scale=0.7]{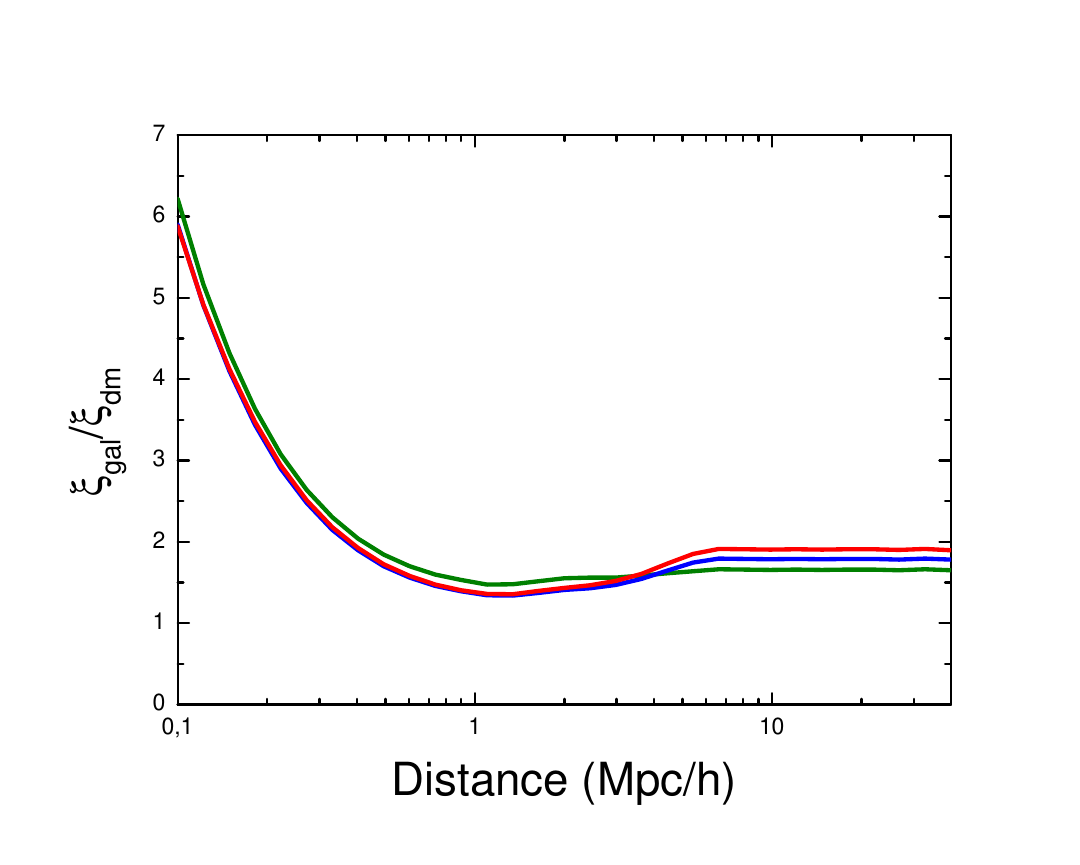}
\caption{Effect of the segregation parameter. Left panel is $\xi_{gal}$ and right panel $\xi_{gal}/\xi_{dm}$. For ${\cal S}=1$ (green line) the red and blue galaxies are completely separated: red galaxies live in node-like haloes and blue galaxies in filament-like haloes. For ${\cal S}=0$ (red line) the red and blue population are perfectly mixed in node- and filament-like haloes. For ${\cal S}=0.5$ (blue line) an intermediate situation is presented:  red galaxies are more abundant in node like haloes and blue galaxies in filament-like haloes, but some degree of mixing between these two population is allowed.  In the left panel, the black-dashed line represents $\xi_{dm}$.}
\label{plot_segregation}
\end{figure}

In Fig. \ref{plot_segregation} we explore the effect of having  the two galaxy populations  mixed or segregated. For two population of galaxies completely segregated the segregation index is ${\cal S}=1$ (green line); when the two population are perfectly mixed this segregation index is ${\cal S}=0$ (red line) (see Eq. \ref{segregation} for definition of $\cal S$). Intermediate cases $1>{\cal S}>0$ ( ${\cal S}=0.5$ for blue line) describe a certain degree of mixing between the two populations, but always with red galaxies being the most abundant population in node-like haloes and blue galaxies more abundant one in filament-like haloes. The galaxy correlation function and $\xi_{gal}/\xi_{dm}$ are plotted in the left and right panel respectively. In the left panel, the black-dashed lines is the dark matter correlation function, $\xi_{dm}$
We observe that at large scales, the more mixed the two population of galaxies are, the higher is the correlation function. This can be understood from the fact the red galaxies are  more biased (see Fig. \ref{plots_hods}).  If the two population are completely mixed (${\cal S}=0$) all haloes will contain red  galaxies and therefore, in total, there  will be more red galaxies than if only the node-like haloes could host this red population (as it happens with ${\cal S}=1$). Therefore  with increasing ${\cal S}$ the relative importance of the more biased galaxy population decreases, thus decreasing the total correlation function. However this effect it is very small in the total correlation function.

\section{Discussion \& Conclusions}
In the classic halo model the host halo mass is the only variable specifying  halo and  galaxy properties. This model has been remarkably  successful at describing the first moment statistics of the clustering of galaxies. However, environment  must play an 
important role in the process of galaxy formation, the most striking observational evidence being that clusters today have a  much higher fraction of early type galaxies than is found in the field. In addition, when looking at statistics beyond the  two-point  density correlation function, there are indications that   the simplest halo-model may be incomplete.

When looking at the  two-point galaxy correlation function for current, low-redshift surveys,  the data can be well modelled  without such an additional dependence on local density (beyond that introduced by the  host halo mass, see e.g., \citep{Skibba09}). However, the statistical power and redshift coverage of  forthcoming surveys implies that an extended modelling may be needed.

In this work we have presented a natural extension of the halo model that allows us to introduce an environmental dependence based on whether a halo lives in a high-mass density region, namely a node-like region, or in a low-mass region, namely a filament-like region. At the level of dark matter, the secondary variable (the environment-dependent variable beyond the shape of the mass function) is the concentration of the density profile of the haloes: haloes which live in nodes regions may have a different concentration than haloes which live in filaments regions. According to this idea we present the dark matter correlation function, $\xi_{dm}$, of this new extended halo model: Eqs. \ref{xidm1}- \ref{xidm3}. In the classic halo model the correlation function is usually split in two terms: the one-halo and the two-halo term. In our extended halo model we have 3 terms: the one-halo term and 2 two-halo terms, depending whether the two particles belong to haloes of the same or different type (see Fig. \ref{our_terms} for a graphical example). We have explored the contribution of these terms to the dark matter correlation function and we have found that the contribution of particles that belong to haloes of the same and different kind is similar (see right panel of Fig. \ref{plot_concentration}). We have also seen that according to the values of table \ref{volume_table}, the contribution of the node-like haloes starts to be dominant at small scales and is subdominant at large scales (see left panel of Fig. \ref{plot_concentration}). We also have analysed how different values of the concentration affect  the correlation function. At the level of dark matter, changing the concentration parameter between 2 and 10 only affects moderately the correlation function at small scales (see left panel of Fig. \ref{concentration_ehm}).

We also have extended our environment-dependent halo model to galaxies, computing the galaxy correlation function, $\xi_{gal}$: Eqs. \ref{xigal1}-\ref{xigal3}. This has been done by modulating the Halo Occupation Distribution recipe with the environment. In the galaxy correlation function, the HOD is our secondary variable rather than halo concentration. This means that according to our model, node-like haloes  host galaxies in a different way than filament-like galaxies. In this paper we use a simple model for the HOD of only 3 variables: the minimum mass of a halo to host its first (central) galaxy, $M_{min}$; the mass of the halo to have on average its first satellite galaxy, $M_1$  and the way how satellite galaxies increase with the halo mass, $\alpha$.  Thus we have two different HOD, one for red-galaxies and another for blue-galaxies. Here we have chosen the colour as an example, although another property besides the colour, like star formation rate or metallicity may be used instead.  Of course more sophisticated HOD could also be used, but starting with a simple prescription helps gaining physical insight. We analyse how changing the different parameters of the HOD affects to the galaxy correlation function (see Fig. \ref{mmin_plot}-\ref{alpha_plot_ehm}). We see that, even when we change the parameters of the HOD of the minority population (in this case the node-like haloes), there is a considerable change in the $\xi_{gal}$, especially for the parameters $M_{1}$ and $\alpha$, i.e., for the satellite population. Therefore changing the HOD of only a small fraction of haloes affects considerably to the total galaxy correlation function. We also have explored how the mixing or segregation between the two population of red and blue galaxies may affect to $\xi_{gal}$. We expect that the size of  this effect will depend on the HOD of each one of these populations: the more different are the HOD of red and blue galaxies, the more the effect of segregation.

It is reasonable to expect the dependence of galaxy properties and clustering on environment to be very complex in details. But, being a “second-order” effect,  it is likely that a simplified description that captures the  main trends will be all that in needed in practice.  The model presented here is simplified in several ways: 1) the environment is only divided into nodes, filaments, and voids  and it does not have a continuous distribution; 2)  that dark halo clustering only depends on mass not on environment.  The formalism introduced here  allows one to implement 2)  straightforwardly.

A natural extension of this model is to make a continuous dependence on the environment, instead of splitting it in nodes, filaments and voids. However this extension presents several issues. First of all we would need a continuous dependence on the environment of the mass function and also of the HOD, which is not yet available.  Secondly, nowadays N-body simulations start to provide information of haloes according a discrete number of environments: this discretization could be taken as a first approximation to the continuous distribution. While this avenue is worth pursuing, introducing a more complex model based on a continuous environment dependence goes beyond the scope of this paper.

We envision  that the extension of the halo model presented here  will be useful for future analysis of large scale structure of the Universe, especially analyses that account for the physical properties of galaxies. Current photometric and spectroscopic large-scale surveys  are beginning to gather not only the position of galaxies but also some of their physical properties, like colour, star formation, age or metallicity, and  with future surveys the accuracy of these measurements will increase. We expect that the physical properties of galaxies  depend on the environment, making these datasets  the suitable ground to apply   the extended halo model.

The halo model presented here uses analytic expressions for the mass function, halo density profile, and also it uses a given HOD for a certain magnitude-selected galaxies. We understand that these analytic functions may be inappropriate for comparison with data. In the present paper we wanted to have an expression for the dark matter non-linear power spectrum, that is fast and easy  to compute and a reasonable approximation to reality; we however were interested in the relative effect of including environment dependence and less concerned with  absolute accuracy of the fit. In a forthcoming paper we are planning to apply this  technique to fit observational data, in this case we would have to further refine our prescription to compute $\xi_{dm}$ (see \cite{smithpeacock} for a numerical approach to the halo model description). This is still work in progress but we believe it goes beyond the scope of the present paper.

 The environmental halo model presented here  is especially suited to model the marked (or weighed) correlation formalism \citep{harker06,Skibba06},  consisting  on weighting each galaxy according to a physical property and removing the dependence of clustering on  local number-counts. Here we have laid the foundations for modelling a survey's marked correlation,  which treatment  will be presented in a forthcoming work.

\section{Acknowledgements}
HGM is supported by CSIC JAE grant.  LV and RJ are supported by MICCIN grant AYA2008-0353. LV is supported by  FP7-IDEAS-Phys.LSS 240117, FP7-PEOPLE-2007-4-3-IRGn202182. We thank B. Reid and R. Sheth for discussions.
\bibliographystyle{mn2e}

\section{Appendix A. The halo model}

In this section we review the basics of the standard halo model. This background material  set up the stage for motivating and introducing our extension of the standard halo model and will define symbols and nomenclature used.

The halo model  pioneered in \citet{peacock_smith,ma_fry,seljak2000} then thoroughly reviewed in  \citet{cooray_sheth},  assumes that all the mass in the Universe is embedded into units, which are called dark matter haloes or simply haloes. These haloes  are small compared to the typical distance between them (non-linear evolution makes the evolved Universe dominated by voids). For this reason, the clustering properties  of the mass density field, $\delta({\bf x})$, on small scales are determined by the spatial distribution inside the dark matter haloes, and the way they are organised in the space is not important. On the other hand the statistics of the large-scale distribution is not affected by the  matter distribution inside haloes but only by their spatial distribution. 
In this work we assume all the time halo exclusion, treating haloes as hard spheres (sharp cutoff at the virial radius). In the case that two particles that belong to different haloes, are separated by less distance that the sum of the virial radii of their host halos, we avoid counting them in the computation of the correlation function.

\subsection{Mass function} 

The number density of collapsed haloes of mass $m$ per unit of mass at a given redshift $z$, $n(m,z)$, can be computed using the Press-Schechter formalism \citep{ps} identifying the present collapsed haloes with the peaks of an initially Gaussian field,
\begin{equation}
 n(m,z)=\frac{2\bar\rho_m}{m\sigma(m)} f(\nu)\nu \left|\frac{d\sigma(m)}{dm}\right|
\label{mass_function}
\end{equation}
where $\bar\rho_m$ is the mean density of matter in the Universe\footnote{According to a $\Lambda CDM$ flat universe ($\Omega_m=0.27$, $\Omega_\Lambda=0.73$ and $h=0.7$) the matter density is given by $\bar\rho_m=\frac{3\Omega^0_m H_0^2}{8\pi G}$ where $H_0\equiv100h$. Using the fiducial cosmology assumed here this value is $\bar\rho_m=7.438\times10^{10} M_\odot/h\left(\mbox{Mpc}/h\right)^{-3}$}, $\sigma(m)$ is the $rms$ of the power spectrum linearly extrapolated to $z=0$ filtered with a top-hat sphere of mass $m$ and $\nu\equiv\delta^2_{sc}(z)/\sigma^2(m)$. Here, $\delta_{sc}(z)$ is the linearly extrapolated critical density required for spherical collapse at $z$, and is given by $\delta_{sc}(z)=1.686/D(z)$, where $D(z)$ is the growth factor. Following this formalism, structures with a linearly evolved density fluctuation higher than this threshold value, will collapse. 
The ST approach  \citep{st}, based on ellipsoidal collapse, predicts a mass function of the form,
\begin{equation}
 f(\nu)\nu=A(p) \left(1+(q\nu)^{-p}\right)\left(\frac{q\nu}{2\pi}\right)^{1/2}\exp\left(-q\nu/2\right)
\label{ST_mass_function}
\end{equation}
with $p=0.3$ and $q=0.75$ and the normalisation factor $A(p)=(1+2^{-p}\Gamma(1/2-p)/\Gamma(1/2))^{-1}$.
Note that by construction the matter density is given by,
\begin{equation}
 \bar\rho_m=\int_0^\infty dm\, n(m,z) m\,.
\end{equation}
In order to introduce an environmental dependence in the mass function, one could think of rescaling Eq. \ref{mass_function} to account the differences of densities. However \cite{mo_white} noted that dark halo abundance in dense and underdense regions do not differ by just a factor like this. A more complex model was proposed by \cite{AbbasSheth1},
\begin{equation}
n(m;M_i,V_i)=\left[1+b(m)\delta_i\right]n(m)
\label{mass_function2}
\end{equation}
where, $n(m; M_i,V_i)$ is the mass function of a region of volume $V_i$ which contains a mass $M_i$, $b(m)$ is the bias of a halo of mass $m$ (see Eq. \ref{bias} for details), and $\delta_i$ is defined as $M_i/V_i\equiv\bar\rho_m(1+\delta_i)$
\begin{figure}
 \centering
\includegraphics[scale=0.7]{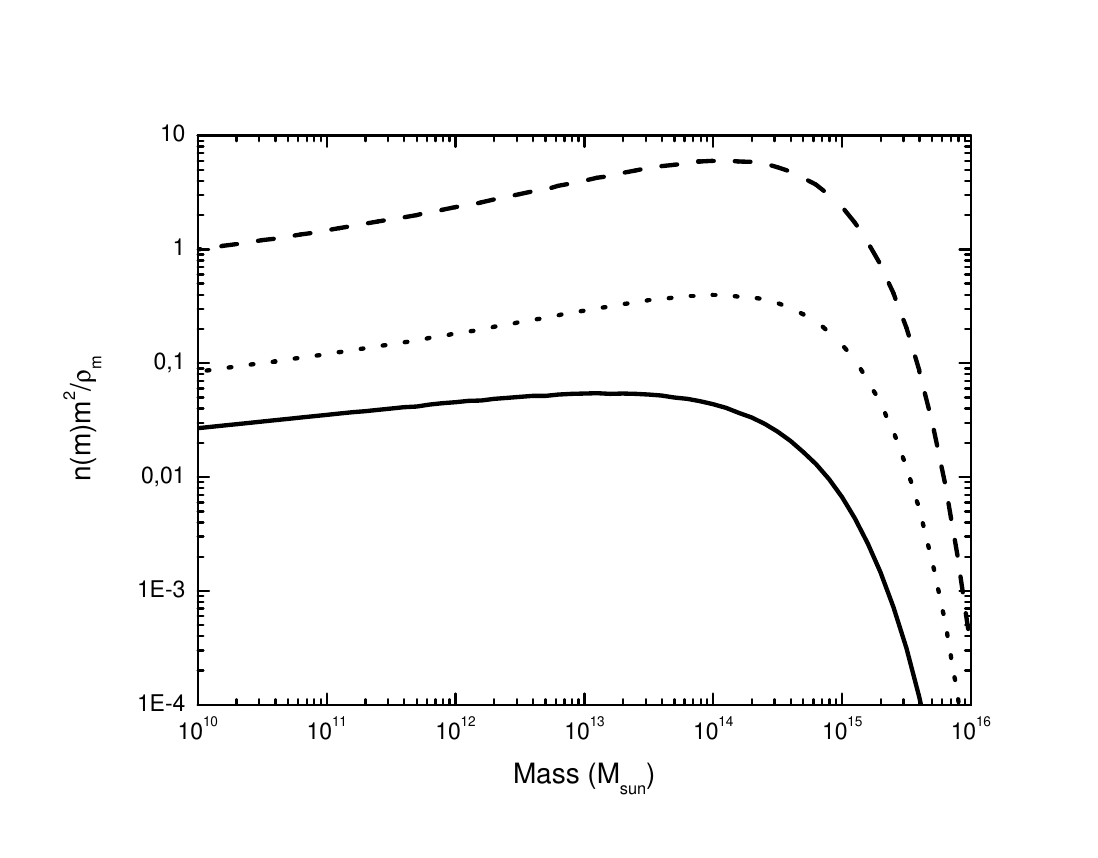}
\includegraphics[scale=0.7]{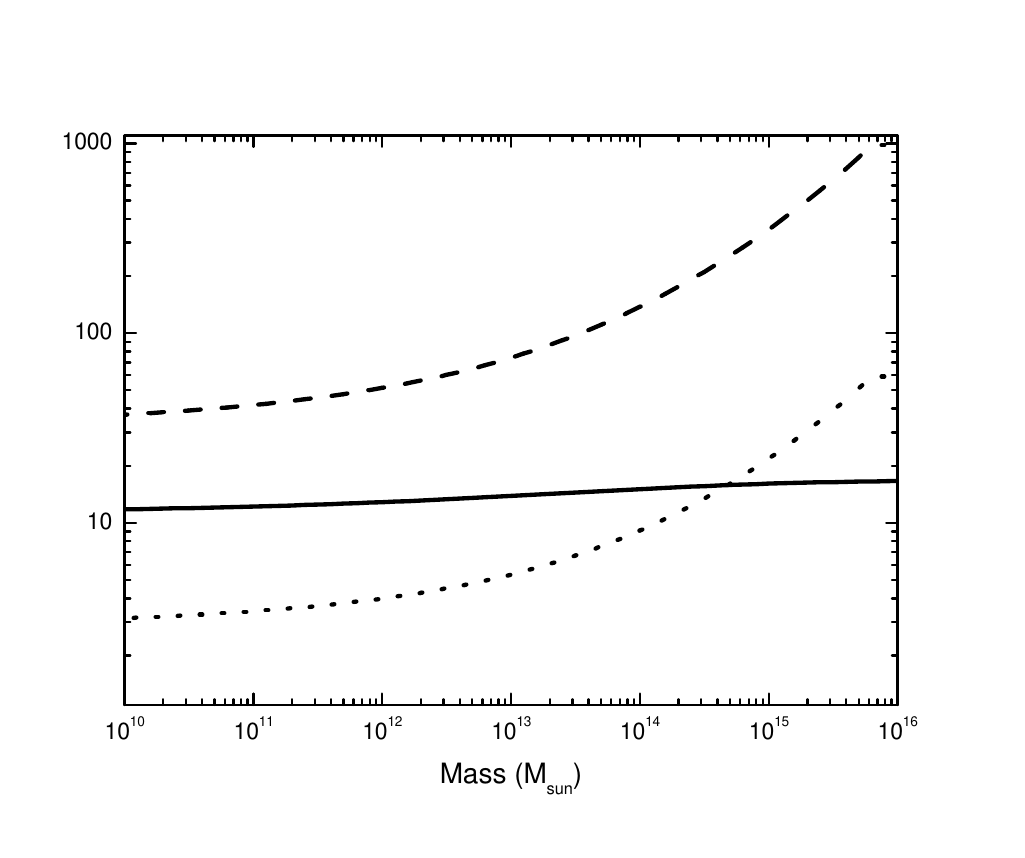}
\caption{Left panel: global mass function (solid line), mass function in node-environment (dashed line) and mass function in filament environments (dotted line). Right panel: Ratio of the mass functions of the left panel. $n_{nod}(m)/n(m)$ (dashed line), $n_{fil}(m)/n(m)$ (dotted line) and $n_{nod}(m)/n_{fil}(m)$ (solid line).}
\label{mass_function_plot}
\end{figure}
In Fig. \ref{mass_function_plot} (left panel) we have plotted the global mass function (Eq. \ref{mass_function}) (solid line), a high-density environment mass function (dashed line) and a medium-density environment mass function (dotted line) using in all cases the ST mass function (Eq. \ref{ST_mass_function}). For these two last mass functions we have used respectively the relative density parameters of table \ref{volume_table} corresponding to nodes and filaments. On the right panel we show the ratios of these three different mass functions: $n_{nod}(m)/n(m)$ (dashed line), $n_{fil}(m)/n(m)$ (dotted line) and $n_{nod}(m)/n_{fil}(m)$ (solid line). We see that the ratio of the number of massive to low mass haloes is larger in dense regions (nodes) than in less dense regions (filaments) since the slope of the solid curve of the right panel of Fig. \ref{mass_function_plot} is positive.

\subsection{Bias}

For large separations   the bias  is the  relation between the correlation function of two dark-matter haloes of masses $m'$ and $m''$ separated by a distance $r$ at a given redshift $z$, $\xi_{hh}(r,z,m',m'')$, and the underlying dark matter linear power spectrum, $\xi_{lin}(r,z)$.  While computing the exact form of $\xi_{hh}(r,z,m',m'')$ is a somewhat delicate matter, excellent results are obtained by using
\begin{equation} 
 \xi_{hh}(r,z,m',m'')\simeq b(m',z)b(m'',z)\xi_{lin}(r,z)
\label{xi_hh}
\end{equation}
where $b(m,z)$ denotes the large-scale linear bias factor  for haloes of mass $m$ at redshift $z$. While using $\xi_{lin}$ as a proxy for $\xi_{hh}$ is strictly incorrect because there may be mildly non-linear contributions and because  at small separations haloes are spatially exclusive, on these scales the signal is dominated by the one-halo term. While the fit to N-body simulations can be further  improved by using e.g.,  higher-order perturbation-theory prediction instead of $\xi_{lin}$ e.g., \cite{SmithShethScocc}, we will not pursue this here as it is beyond the scope of the present paper. 

To be consistent we should use a bias prescription  derived from the extended Press-Schechter formalism using the peak background split. To the lowest order the bias is,
\begin{equation}
b(m,z)=1-\left.\frac{\partial \ln[n(m,z')]}{\partial \delta_{sc}(z')}\right|_{\delta_{sc}(z)} 
\end{equation}
The bias of an object $m$ at time $z$ is given, at lowest order of $\delta$, by \citep{cole_kaiser,mo_white}
\begin{equation}
 b(m,z)=1+\frac{1}{D(z)}\left[q\frac{\delta_{sc}(z)}{\sigma^2(m)}-\frac{1}{\delta_{sc}(z)}\right]
 \label{bias}
\end{equation}
where, $D(z)$ is the linear growth factor, $\delta_{sc}(z)$ is the critical threshold for collapse and is given by $\delta_{sc}\simeq1.686/D(z)$, $\sigma(m)$ is the $rms$ of the power spectrum linearly extrapolated at $z=0$ filtered with a top-hat sphere of mass $m$ and $q$ is the parameter introduced in the Sheth \& Tormen mass function (see Eq. \ref{ST_mass_function}).
This formula has been confirmed by N-body simulations giving an excellent agreement.
Although the complete formula should have the extra term, $2p/(\delta_{sc}(z)D(z))\left(1+\left(\frac{q\delta_{sc}^2(z)}{\sigma(m)^2}\right)^p\right)^{-1}$, we have checked that the effect of this term is about $1\%$ and can be safely neglected for our application.

\subsection{Density profile}

According to our definition, a halo is a set of particles that forms a gravitationally bound and thermodynamically stable system. Therefore, we consider that this system satisfies the virial theorem, i.e. a halo is virialised by definition. From the spherical collapse model \citep{gunn_gott}, a halo is considered to be  formed and virialised when its density reaches a certain threshold value: $\rho_{vir}$. 
Here we will adopt $\bar\rho_{vir}=18\pi^2\bar\rho_m$, which is the typical value that can be found using the spherical collapse model. The corresponding virial radius is then,
\begin{equation}
 r_{vir}(m)=\frac{1}{2\pi}\left(\frac{m}{3\bar\rho_m}\right)^{1/3}
\end{equation}
In the halo model, this will be the size of an halo of mass $m$.

On the other hand, cosmological simulations have shown that the density profile inside isolated haloes follows a universal profile given by \citep{nfw96,nfw97},
\begin{equation}
 \rho(r|r_s,\rho_s)=\frac{\rho_s}{(r/r_s)(1+r/r_s)^2}
\end{equation}
where $r_s$ is the scale radius of the halo and $\rho_s/4$ is the halo density at scale radius. However it is more natural to work with the concentration $c$, and the mass of the halo $m$, instead of $\rho_s$ and $r_s$ when we describe the halo profile. These variables are defined as,
\begin{equation}
 c\equiv r_{vir} /r_s 
\end{equation}
and the mass of the halo has to satisfy 
\begin{equation}
 m=\int_0^{r_{vir}(m)}dr\,4\pi r^2\rho(r|r_s,\rho_s)\,.
 \label{mass_def}
\end{equation}
From this last equation, we can write,
\begin{equation}
 \rho_s=\frac{m}{4\pi r_s^3}\left[\ln(1+c)-\frac{c}{1+c}\right]^{-1}\,.
\end{equation}

We prefer to use normalised profile of a halo $u$ as a function of distance from its centre for given halo virial mass and concentration parameter,   defined by,
\begin{equation}
 u(r|m,c)\equiv\frac{\rho(r|m,c)}{m}
\label{nor_profile}
\end{equation}
which  satisfies the condition 
\begin{equation}
1=\int_0^{r_{vir}(m)}d^3{\bf x}\, u(x|m,c)\,.
\end{equation}

\subsection{Concentration}

In principle the halo  density profile depends on two independent parameters: either $\rho_s$ and $r_s$, or $c$ and $m$. However N-body  simulations indicate that there is a  relation between the concentration and the mass. \cite{seljak2000} propose
\begin{equation}
 c(m,z)=\frac{9}{1+z}\left[\frac{m}{m_\star}\right]^{-0.2}
 \label{concentration}
\end{equation}
where $m_\star$ is the mass of a typical collapse halo at $z=0$. There are many other parametrisations of the relation between the concentration and the mass  (e.g., \cite{neto07, zhao09}). However we do not expect that  the results of this work  to depend significantly on this choice. Unless otherwise stated, for the present application we thus adopt Eq. \ref{concentration} for the relation between concentration and mass. 

This relation goes in the direction one may have expected: less massive  haloes on average form earlier, when the Universe is more concentrated and therefore have a higher concentration than more massive ones. One should bear in mind  however that there is a large dispersion around this mean relation, which may correspond to a "hidden parameter" such as local environmental effects, tidal effects, mergers etc.

\subsection{Two-point correlation function}
\begin{figure}
 \centering
\includegraphics[scale=0.3]{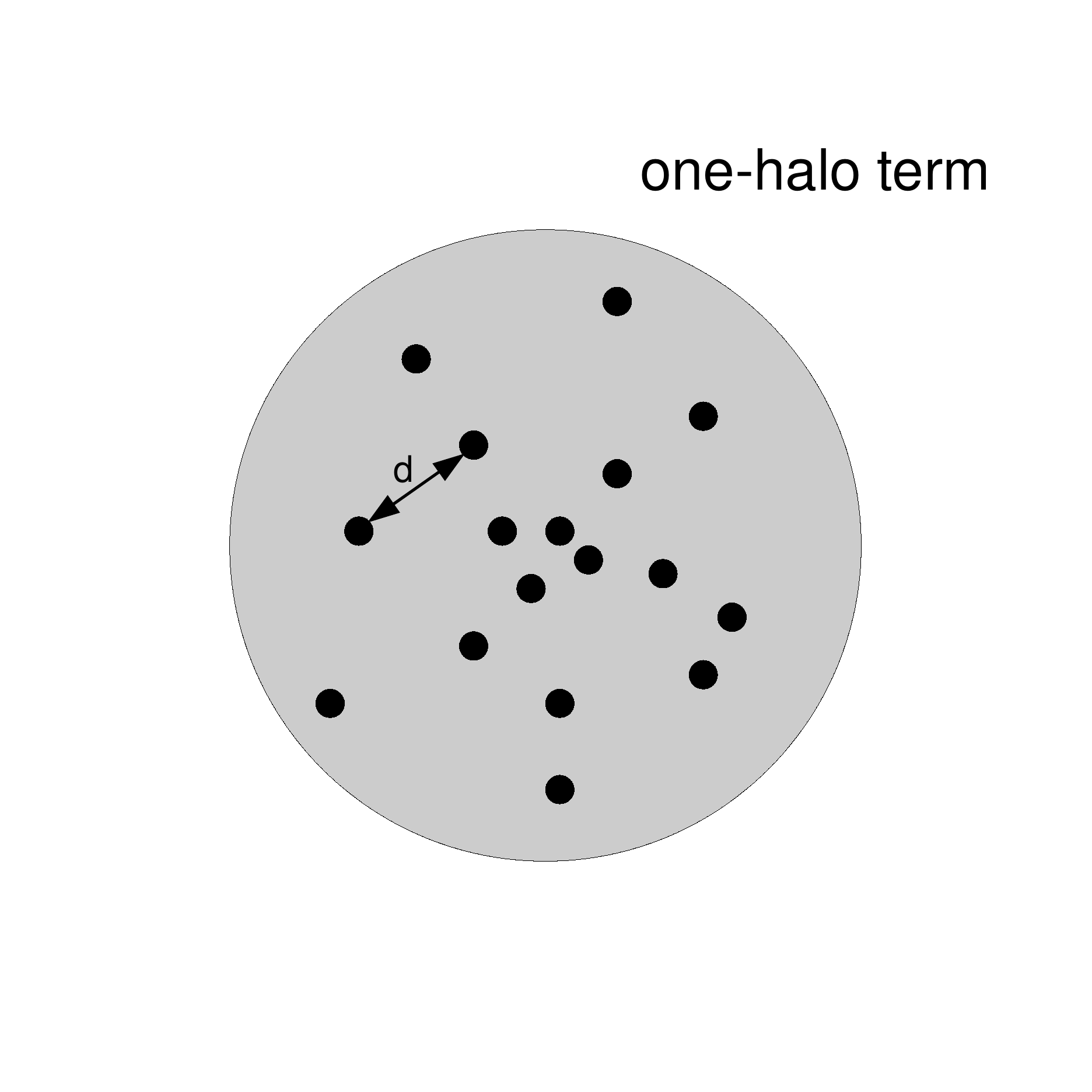}
\includegraphics[scale=0.3]{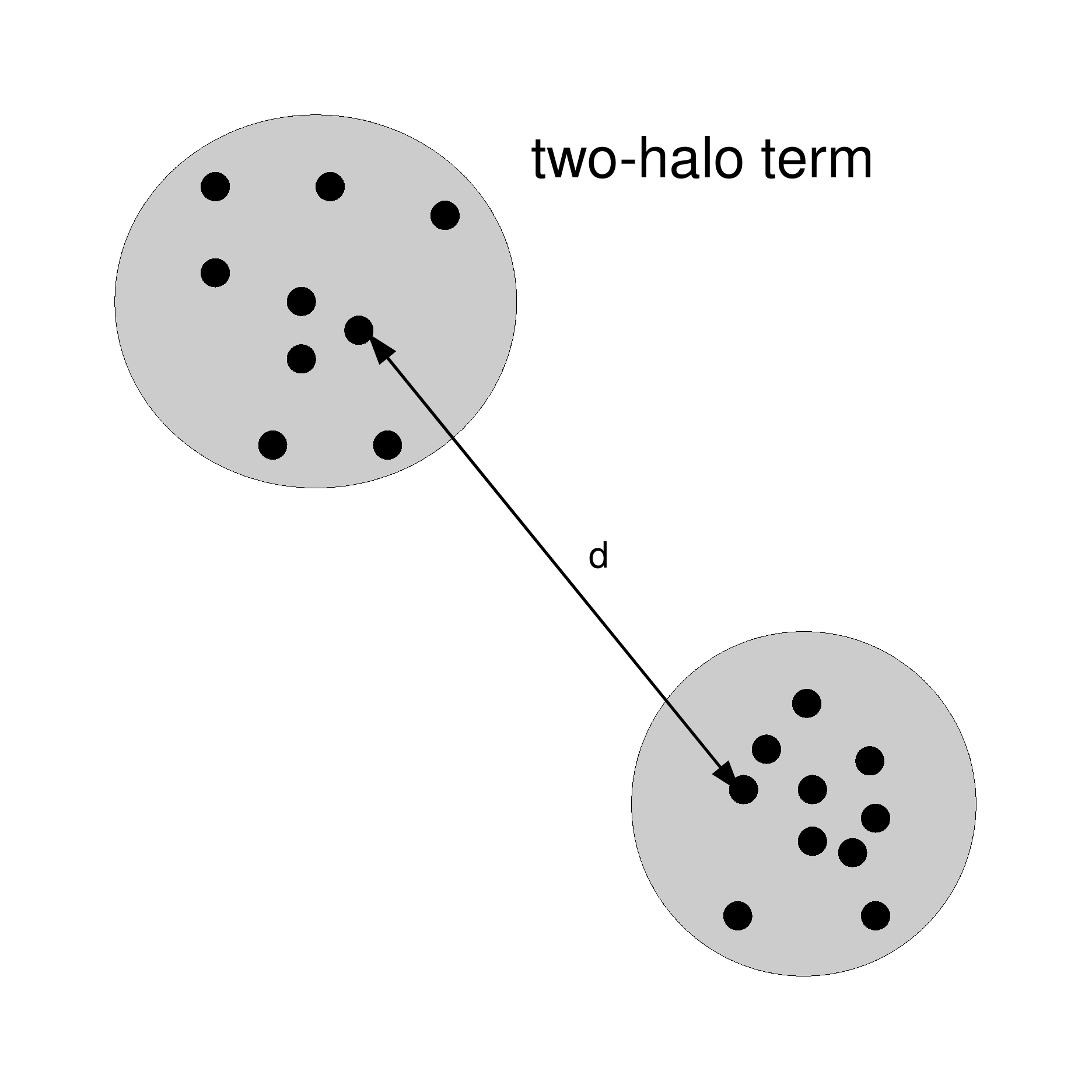}
\caption{In the halo model, the correlation function is split in two terms. The one-halo term (left) describes the clustering of particles inside the same halo. On the other hand, the two-halo term (right) involves only particles of different haloes.}
\label{terms}
\end{figure}

In the halo model, the two-point correlation function of dark matter particles  contained in haloes is defined as,
\begin{equation}
 \xi_{dm}({\bf r})\equiv<\delta({\bf x})\delta({\bf x + r})>
\end{equation}
where $\delta({\bf x})\equiv\rho({\bf x})/\bar\rho_m-1$. This function can be split into 2 terms, depending on whether the two particles at distance $\bf r$ belong or not to the same halo:
\begin{equation}
 \xi_{dm}({\bf r})=\xi_{dm}^{1h}({\bf r})+\xi_{dm}^{2h}({\bf r}).
\end{equation}
$\xi_{dm}^{1h}(\bf r)$ is called the one-halo term and accounts for particles in the same halo;  $\xi_{dm}^{2h}(\bf r)$ is called the two-halo term and accounts for particles  belonging to different haloes. Therefore the properties of the mass density on small scales are described by $\xi_{dm}^{1h}(r)$, whereas on large scales are given by $\xi_{dm}^{2h}(r)$.
A graphical description is shown in Fig. \ref{terms}.

According to the halo model formalism, these terms are,
\begin{eqnarray}
 \xi_{dm}^{1h}({\bf r},z)&=&\int dm\, \frac{m^2\, n(m,z)}{\bar\rho_m^2(z)} \int_V d^3 {\bf x}\,u({\bf x}|m)\,u(|{\bf x+ r}||m)\label{1haloterm}\\
\xi_{dm}^{2h}({\bf r},z) &=& \int dm' \, \frac{m'\, n(m',z)}{\bar\rho_m(z)} \int dm'' \, \frac{m''\, n(m'',z)}{\bar\rho_m(z)}   \int_V d^3{\bf x'}\, u(x'|m')  \int_V d^3{\bf x''}\, u( x''|m'')\xi_{hh}(|{\bf x' - x''+ r}|,z,m', m'')\label{2haloterm}
\end{eqnarray}
The integration $\int_Vd^3{\bf x}$ means over all haloes' volume, and $\int dm$ runs over all mass range. For computational reasons we have to adopt some limits in the mass integral. Setting $m_{max}=10^{15}M_\odot$ is enough as far as the value of the integral do not change if we increase this limit. If we set the minimum mass limit to $10^9 M_\odot$ it is also enough for the 1-halo term, because the less massive haloes contribute at very small distances (less than 0.1 Mpc/h). However the contribution of low mass haloes to the 2-halo term its not negligible. In order to compute these integrals we use the method described in \cite{yoo}  which consists in breaking the 2-halo integral in two parts and approximate the low mass halos as points without inner structure,
\begin{equation}
\int_0^\infty dm \frac{m n(m,z)}{\bar\rho_m}b(m,z)\int_V d^3{\bf x}\, u(x|m)\simeq\int_{m_{th}}^\infty dm \frac{m n(m,z)}{\bar\rho_m}b(m,z)\int_V d^3{\bf x}\, u(x|m)+\left[1-\int_{m_{th}}^\infty \frac{m n(m,z)}{\bar\rho_m}b(m,z)\right]  
\end{equation}
where we have used the approximation of Eq. \ref{xi_hh} to explicit the mass dependence of $\xi_{hh}$.
In particular we have set $m_{th}$ to $10^9 M_\odot$ and we have checked that reducing this value do not affect the result of the integral

\begin{figure}
\centering 
\includegraphics[scale=1.0]{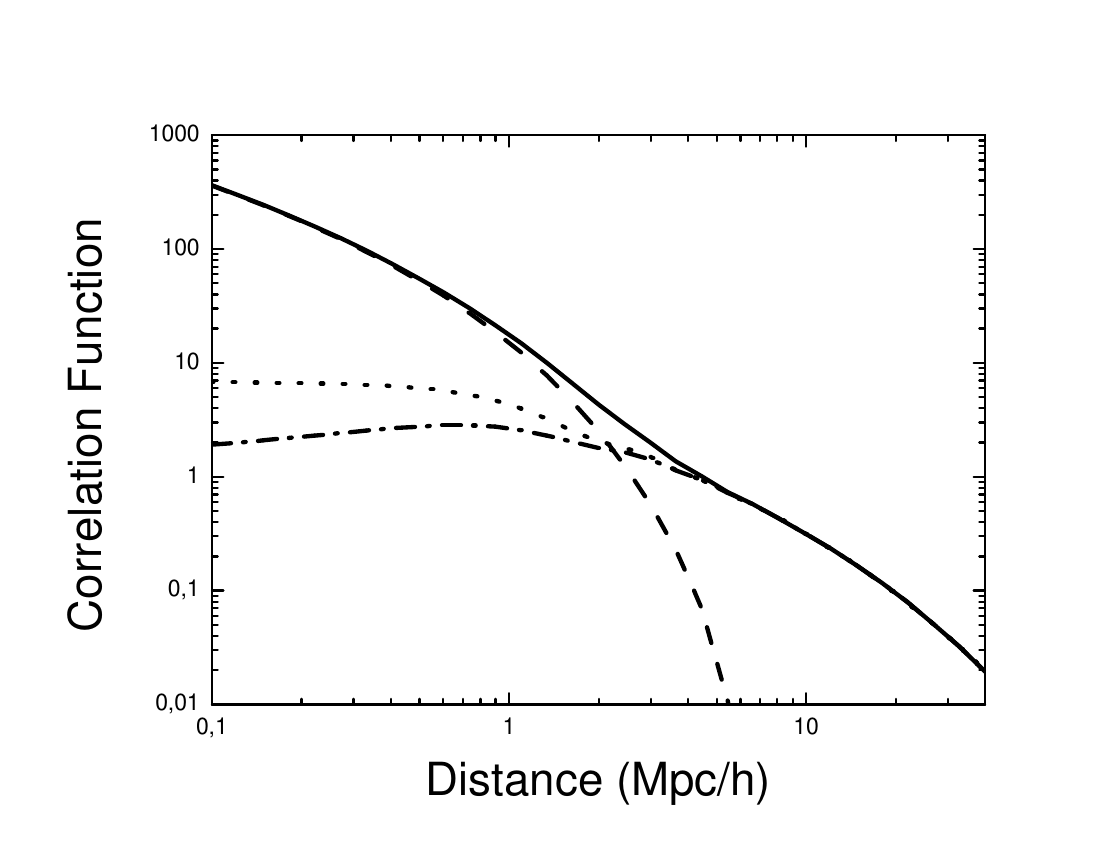}
\caption{Total dark matter correlation function $\xi_{dm}$ (solid line), one-halo term $\xi_{dm}^{1h}$ (dashed line), two-halo term $\xi_{dm}^{2h}$ (dot-dashed line)  and linear power spectrum $\xi_{lin}$ (dotted line).}
\label{halo_terms}
\end{figure}

In Fig. \ref{halo_terms} we can see the contribution of these two terms. $\xi^{1h}_{dm}(\bf r)$ dominates at scales typically smaller than the virial radius, whereas $\xi^{2h}_{dm}(\bf r)$ does it for larger scales following the shape of $\xi_{lin}$ slightly shifted by the effect of the bias (see Eq. \ref{xi_hh}).

\subsection{Halo occupation distribution }\label{section_hod_chm}

The halo model provides a framework to also model galaxy clustering: the complicated galaxy formation physics would determine how many galaxies form in a halo and their sampling of the dark matter  halo profile. Thus, the shape of the one-halo term would be modified  accordingly.

Typically the HOD  assume a centre-satellite distribution of galaxies inside each halo. This means that a galaxy is placed  at the centre of the halo and may be  surrounded by satellite galaxies distributed according to some statistics. The average number of galaxies that lie in a halo of mass $m$ is  (e.g., \cite{sheth05}),
\begin{equation}
g_{(1)}(m)\equiv\sum_N N p(N|m)
\end{equation}
where $p(N|m)$ is the probability density of $N$ galaxies are formed in a halo of mass $m$.

We can also define the $n$th factorial moment of the distribution $p(N|m)$ of galaxies in haloes of mass $m$,
\begin{equation}
g_{(n)}(m)=\sum_N N (N-1)\cdots (N-n+1)p(n|m).
\end{equation}
Here we assume that $p(n|m)$ follows a Poisson distribution, although other choices are also possible.  Under this assumption, we can state that for satellite galaxies, 
\begin{equation}
g_{(n)}(m)= (g_{(1)}(m))^n\,.
\end{equation}
In particular we will only be interested in the first moment $g_{(1)}(m)$ (the mean number of galaxies in a halo of mass $m$) and in the second moment $g_{(2)}(m)$ that we will treat as the square power of the former. Therefore hereafter we write $g_{(1)}(m)$ as $g(m)$ to simplify the notation.

The typical way to parametrise the mean number of centre galaxies is to think of the mean number of the central galaxies as a step function,
\begin{equation}
g^{cen}(m)=\left\{
\begin{array}{c} 
1\quad\quad \mbox{if}\quad m> M_{min}\\ 
0 \quad\quad \mbox{if}\quad m< M_{min}\,.
\end{array}\right.
\label{HOD_cen}
\end{equation}
The satellite galaxy distribution can be parametrised as a power law, with the same mass cut as  central galaxies,
\begin{equation}
g^{sat}(m)=\left(\frac{m}{M_{1}}\right)^{\alpha} g^{cen}(m)
\label{HOD_sat}
\end{equation}
where $M_{min}$ sets the minimum mass for a halo to have galaxies; $M_{1}$ is the mass of a halo that on average hosts one satellite galaxy; and $\alpha$ is the power-law slope of the satellite mean occupation function. All these parameters can be tuned to fit  observations and in general depend on the type of galaxy  under consideration.

\begin{figure}
\centering
\includegraphics[scale=0.65]{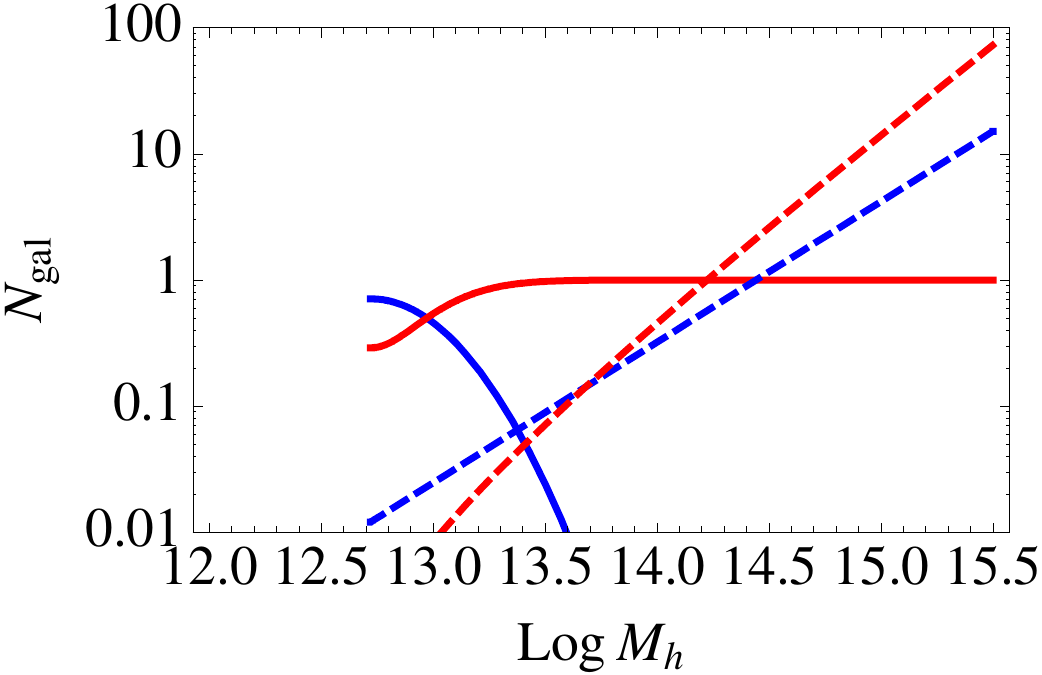}
\includegraphics[scale=0.65]{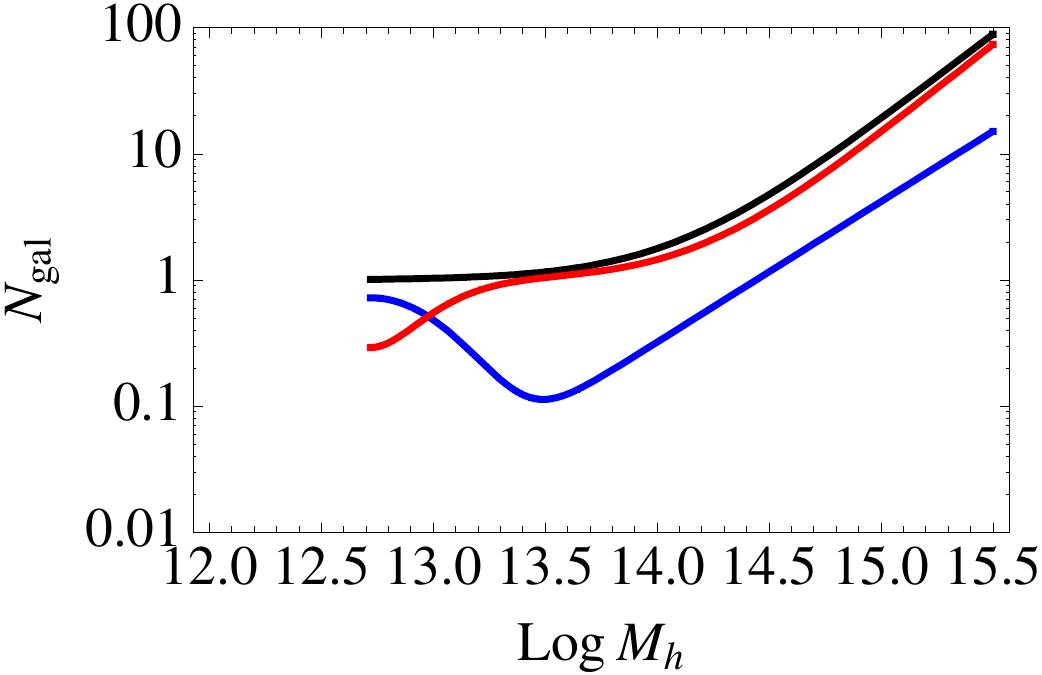}
\caption{HOD with colour dependence according to \citet{zehavi05}. Left plot: Central and satellite galaxy contribution (solid and dashed lines respectively) for red and blue galaxies (red and blue lines respectively). Right panel: Total galaxy number for all galaxies (black line), red galaxies (red line) and blue galaxies (blue line).}
\label{hod_rb}
\end{figure}

\begin{figure}
 \centering
 \includegraphics[scale=0.52]{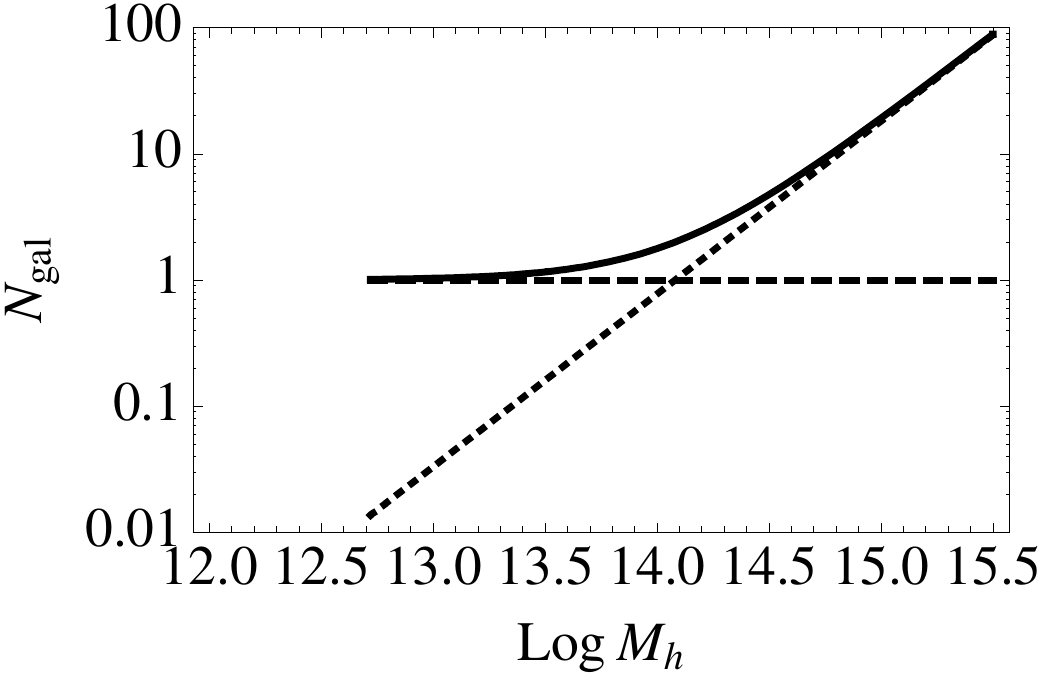}
 \includegraphics[scale=0.52]{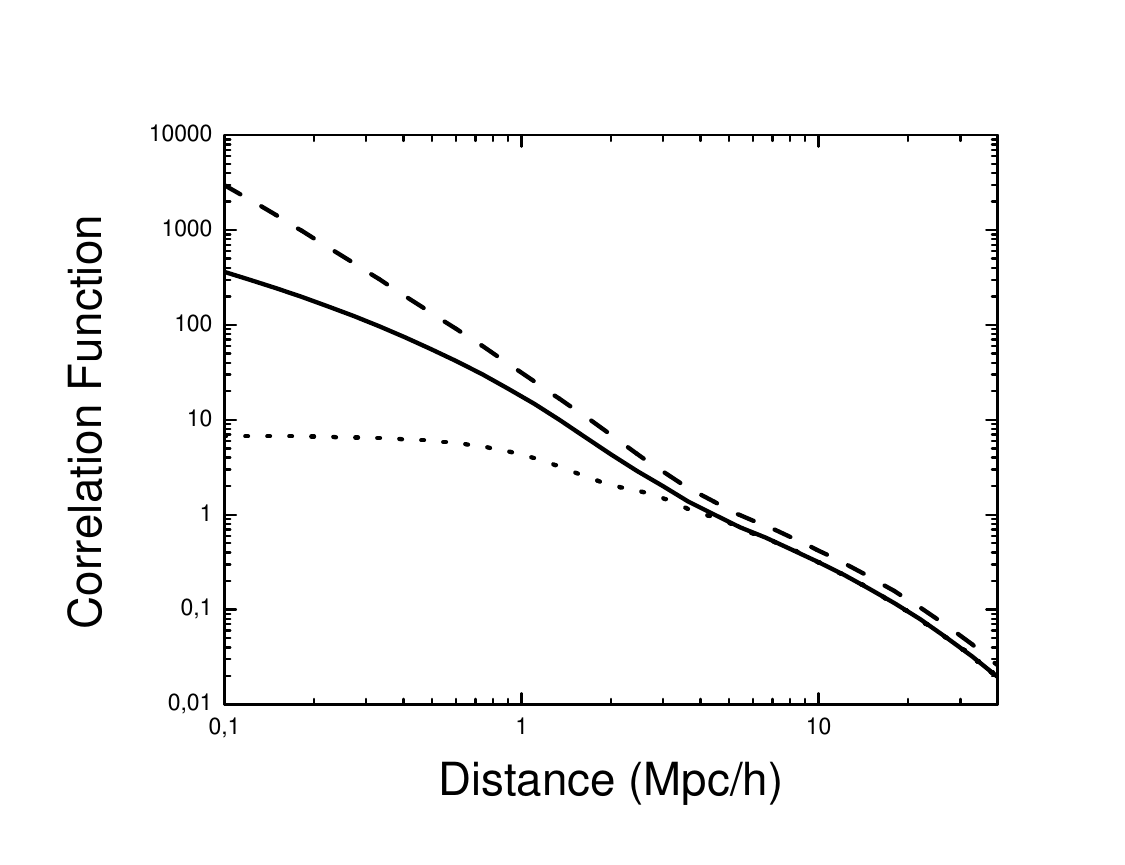}
 \includegraphics[scale=0.52]{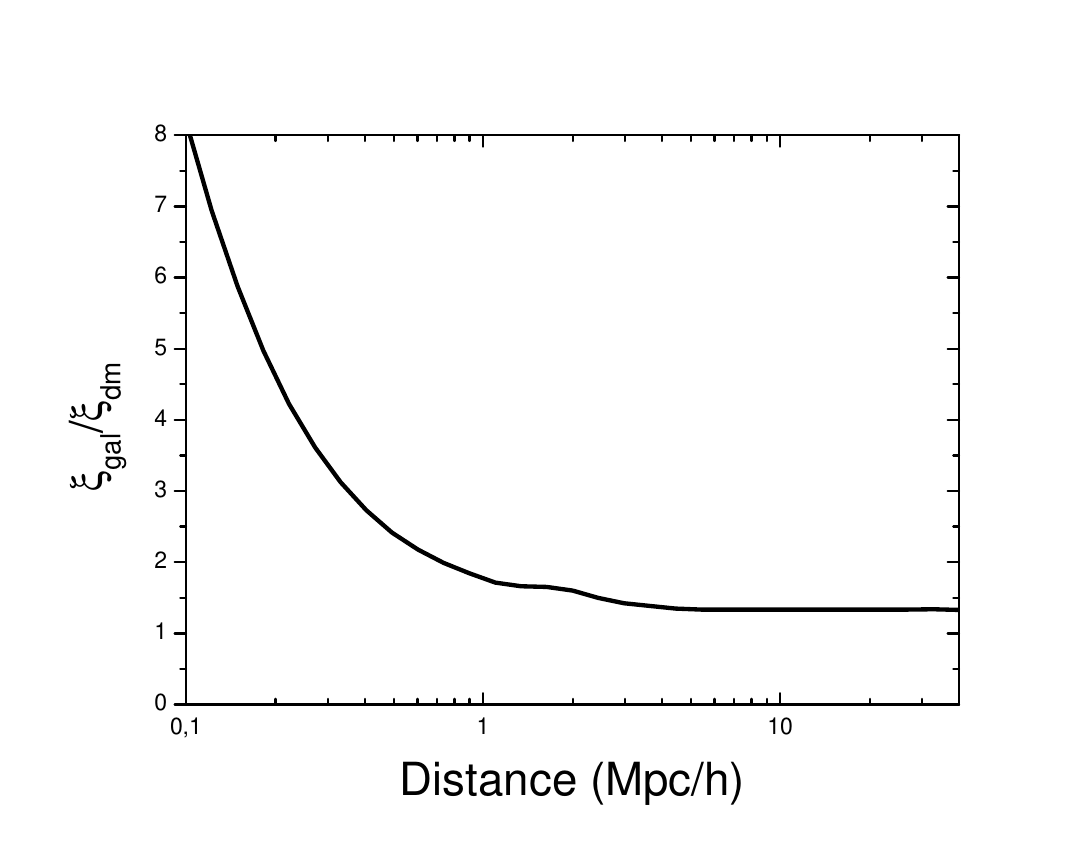}
 \caption{Left panel: HOD for the -$21>M_r$ sample. All galaxies (solid line), central galaxy (dashed line) and satellite galaxies (dotted line). Centre panel: Dark matter correlation function $\xi_{dm}$ (solid line), galaxy correlation function $\xi_{gal}$ (dashed line) according the HOD of the left panel and linear correlation function $\xi_{lin}$ (dotted line). Right panel: ratio between $\xi_{gal}$ and $\xi_{dm}$.}
 \label{hods_galas}
 \end{figure}

In the literature there are several ways to  populate haloes with galaxies depending on their type: LRGs \citep{reid_spergel}, field galaxies \citep{kravtsov04} and many others (e.g., \cite{zehavi05}). On one hand, \cite{kravtsov04} using high-resolution dissipationless simulations find that the parameters of the HOD according to Eqs. \ref{HOD_cen} and \ref{HOD_sat} are $M_{min}\simeq10^{11}M_\odot/h$, $M_1\simeq22 M_{min}$ and $\alpha\simeq1$. On the other hand, \cite{zehavi05} based on SDSS observation state that for a volume-limited sample with magnitude in $r$-band ($M_r$) $<-21$ the values of these parameters are $M_{min}=10^{12.72}M_{\odot}/h$, $M_1=10^{14.08}M_\odot/h$ and $\alpha=1.37$ (see Fig. \ref{hods_galas} left panel), although these values depend strongly on the magnitude selection criteria. Therefore it is clear that the parameters of the HOD depend strongly on the way  galaxies are selected: whether  one deals with a volume-limited sample of  whether a magnitude selection criteria has been applied. It is interesting to note that some authors  (e.g., \cite{zehavi05})  introduce a colour dependence through the HOD: haloes of the same mass host different kind of galaxies, red and blue. Since blue galaxies tend to live in low-density regions (filamentary regions) and red galaxies  tend to live in high-density regions (node regions), this can be though as different HOD parametrization for different kind of environments. In particular, \cite{zehavi05} introduce the fraction of blue galaxies in a halo of mass $m$, $f_b(m)$, as a function of the halo mass and also as a function of galaxy type: central or satellite. 
Since red galaxies are more common in high mass haloes the authors adopt a function for $f_b$ which is decreasing with halo mass.
For central galaxies a log-normal function is adopted:
\begin{equation}
f_b^{cen}(m)=f_0^{cen}\exp\left[-\frac{(\log_{10}m-\log_{10}M_{min})^2}{2{\sigma_M^{cen}}^2}\right]
\end{equation}
whereas for satellite a log-exponential function is used:
\begin{equation}
f_b^{sat}(m)=f_0^{sat}\exp\left(-\frac{\log_{10}m-\log_{10}M_{min}}{\sigma_M^{sat}}\right)\,,
\end{equation}
where $m$ is the mass of the halo, $M_{min}$ is the minimum mass of a halo to host one galaxy, $f_0$ is the value of the fraction for a halo with mass $M_{min}$ and $\sigma_M$ is a parameter that characterise how fast the blue fraction drops.
Also these parameters depend strongly on the magnitude-selection criteria. For $M_r<-21$ they are: $\log_{10}M_{min}=12.72$, $f_0^{cen}=0.71$, $f_0^{sat}=0.88$, $\sigma_M^{cen}=0.30$ and $\sigma_M^{sat}=1.70$.
The average number of blue and red galaxies becomes then,
\begin{eqnarray}
g_{blue}^i(m)&=&f_b(m)^i g^i(m) \nonumber \\
g_{red}^i(m)&=&\left[1-f_b^i(m)\right]g^i(m)
\label{eq:zehavi2}
\end{eqnarray}
where the average number of galaxies of type $i$, $g^i(m)$ is given by Eqs. \ref{HOD_cen} and \ref{HOD_sat} according the values of the parameters used in \cite{zehavi05}. 
In Fig. \ref{hod_rb}  we illustrate the corresponding HOD for Eq. \ref{eq:zehavi2} for  red (red lines) and blue (blue lines) galaxies, and for the total number of galaxies (black line). In the left panel, we show the separate central and satellite contributions: the solid lines are central galaxies and dashed lines satellite galaxies. In the right panel we show the combined effect of the contributions. As stated by the authors, for haloes just above $M_{min}$, blue central galaxies are more common. However above $2M_{min}$, central galaxies are predominantly red. Note that  for blue galaxies there is a minimum in the total number of galaxies that occurs when a halo is too massive to have a blue central galaxy but not massive enough to host a significant population of blue satellite galaxies. As a general trend, we can say that according to this kind of HOD model: {\it i)} the lowest mass haloes have blue central galaxies, {\it ii)} higher mass haloes have red central galaxies but with a significant blue fraction in their satellite population and finally {\it iii)} the highest mass haloes host red central and satellite galaxies. \citet{zehavi05} considered that the two populations are perfectly mixed  and proceeded to compare the HOD  predictions for the projected correlation function with SDSS data. This kind of difference between the HOD of blue and red galaxies motivates us to further explore the environmental dependence of the HODs of haloes of different density regions and in particular to consider that different environment can have different ratios of red and blue galaxies.

Eqs. \ref{1haloterm}, \ref{2haloterm}  can then be modified to describe the clustering of galaxies:

\begin{eqnarray}
\label{classic_gal_hm1}\xi_{gal}^{1h}({\bf  r},z)&=&\int dm\, \frac{n(m,z)}{\bar n_{gal}^2(z)}\left[2\,g^{cen}(m)g^{sat}(m)u(r|m)+\left(g^{sat}(m)\right)^2\int d^3{\bf x}\, u (x|m)\, u(|{\bf x + r}|m)\right]\\
\nonumber\xi_{gal}^{2h}({\bf r},z)&=&\int dm' dm'' \frac{n(m',z)n(m'',z)}{\bar n_{gal}^2(z)}\left[g^{cen}(m')g^{cen}(m'')\xi_{hh}(r,z,m',m'')+2\,g^{cen}(m')g^{sat}(m'')\int d^3{\bf x'}\, u(x'|m'')\right.\times\\
\label{classic_gal_hm2}&\times&\left.\xi_{hh}(|{\bf x' + r}|,z,m',m'')+g^{sat}(m')g^{sat}(m'')\int d^3{\bf x'} d^3{\bf x''}\,  u(x'|m') u(x''|m'')\xi_{hh}(|{\bf x'-x'' + r}|,z,m',m'')\right]
\end{eqnarray}
 where the mean number of galaxies per halo is given by
 \begin{equation}
 \bar n_{gal}(z)=\int dm\, n(m,z) \left(g^{cen}(m)+g^{sat}(m)\right)\,.
 \label{barn}
 \end{equation} 
 
In Fig. \ref{hods_galas} (left panel) the  \cite{zehavi05} HOD is plotted ($M<-21$ sample,  $\log_{10}M_{min}=12.72$, $\log_{10}M_1=14.08$ and $\alpha=1.37$). The dashed line corresponds to the contribution of central galaxies, the dotted line to satellite galaxies and the solid line to the total. In the central panel we show the correlation function for dark matter and galaxies: $\xi_{gal}$ (solid line), $\xi_{dm}$ (dashed line) and $\xi_{lin}$ (dotted line), according the  same HOD. In the right panel, the ratio between $\xi_{gal}$ and $\xi_{dm}$ is plotted. We see that at scales smaller than $\sim1\,\mbox{Mpc}/h$, when typically the one-halo term dominates, the ratio $\xi_{gal}/\xi_{dm}$ is highly scale dependent, whereas at larger scales, then the two-halo terms dominates the ratio $\xi_{gal}/\xi_{dm}$ is scale independent. In particular it appears that at scales $\sim0.1\,\mbox{Mpc}/h$ there is a minimum for $\xi_{gal}/\xi_{dm}$ with a value of 8. At large scales ($>10\, \mbox{Mpc}/h$) $\xi_{gal}/\xi_{dm}$ takes a value around 1.3. 
This is expected from the halo model formalism, where at small scales the 1-halo term dominates whereas at large scales the 2-halo term does it. From Eq. \ref{classic_gal_hm1}-\ref{classic_gal_hm2} it can be seen that at large scales the 2-halo term can be expressed as a function of $\xi_{lin}$ whereas this does not happen for the 1-halo term. This yields us to the point that at large scales the ratio between galaxies and dark matter it is just a constant, whereas at small scales the relation it is much more complicated.

\section {Appendix B}

Here we set the basics of the derivation of Eqs. \ref{xidm1}-\ref{xidm3} of our extended halo model. This derivation is very similar to the one of classic halo model presented for the first time by \cite{sb91}.

We define the overdensity $\delta({\bf x})$ as $\delta({\bf x})\equiv \rho({\bf x})/\bar\rho-1$. Then the dark matter correlation function is,
\begin{equation}
\xi_{dm}({\bf r})\equiv\langle\delta({\bf x})\delta({\bf x+r})\rangle=\frac{1}{\bar\rho^ 2}\langle\rho({\bf x})\rho({\bf x+r})\rangle-1
\label{xidm_app}
\end{equation}
Where $\langle\cdots\rangle$ denotes the ensemble average.
On the other hand, the density field at $\bf x$ is the sum of densities of node- and filament-like regions: $\rho({\bf x})=\rho^{nod}({\bf x})+\rho^{fil}({\bf x})$. We also can write the contribution of the nodes/filaments as the sum every halo of this class:
\begin{equation}
\rho^{j}({\bf x})=\sum_im_i u({\bf x - x_i}| m_i, c_j)
\label{rho_j}
\end{equation}
where the summation takes place over all haloes of type $j$ (nodes or filaments), $m_i$ and $\bf x_i$ are the mass and the position (of the centre) of the $i$th halo and $u$ is the normalised profile defined in Eq. \ref{nor_profile}.
From Eqs. \ref{xidm_app} and \ref{rho_j} it is clear that we can split $\xi_{dm}$ in 3 terms: $\xi_{dm}^{1h}$, $\xi_{dm}^{2h1\eta}$ and $\xi_{dm}^{2h2\eta}$  depending on whether the particles belong to the same halo and depending on the particles belong to the same kind of halo (in the case of the two-halo term). The derivation of each terms to Eqs. \ref{xidm1}-\ref{xidm3} is similar and are based on the introduction of the Dirac delta,
\begin{equation}
\sum_i\rightarrow\int dm\, d^3{\bf x}\, \sum_i \delta^3_D({\bf x-x_i})\delta_D(m-m_i)
\end{equation}
and on the definition of the mass functions in the node- and filament-like regions as,
\begin{equation}
n_j(m)\equiv\frac{1}{V_j}\int_{V_j}\sum_i d^3{\bf x}\,\delta_D^3({\bf x-x_i})\delta_D(m-m_i)
\end{equation}
where the index $j$ denotes the type of environment, either node or filament, and $V_j$ the volume of this environment.

Using these formulae and following the formalism presented by \cite{sb91}, Eqs. \ref{xidm1}-\ref{xidm3} can be obtained after some algebra.

\end{document}